\documentclass[twocolumn,10pt,a4paper]{article}
\usepackage[top=2cm,bottom=2cm,left=1.6cm,right=1.6cm,columnsep=0.7cm]{geometry}
\usepackage{graphicx,amsmath,amssymb,booktabs,siunitx,xcolor}
\usepackage{colortbl}            %
\usepackage[hidelinks]{hyperref}
\usepackage[affil-it]{authblk}
\usepackage{placeins}            %
\usepackage{stfloats}
\usepackage{pdfpages}            %

\newcommand\samethanks[1][\value{footnote}]{\footnotemark[#1]} %
\graphicspath{{fig1/}{fig2/}{fig3/}{fig4/}{fig5/}{appendix/}{appendix1/}{appendix2/}{appendix3/}{tab1/}}

\title{\bfseries\Large Coupled-cluster molecular properties across
the main group that extrapolate beyond training size}

\author[1]{Wenhao He\thanks{These authors contributed equally.}}
\author[3]{Xu Chen\samethanks}
\author[3]{Noah Song\samethanks}
\author[4]{Haowei Xu}
\author[4]{Tim S. Hindges}
\author[1]{Bohan Li}
\author[2]{Zihan Lin}
\author[4]{Yu Yao}
\author[5]{Avetik R. Harutyunyan}
\author[3]{Fang Liu}
\author[3]{Yao Wang}
\author[2]{Hao Tang\thanks{Correspondence: \texttt{haot@mit.edu}}}
\author[1,2,4]{Ju Li\thanks{Correspondence: \texttt{liju@mit.edu}}}

\affil[1]{Center for Computational Science and Engineering,
  Massachusetts Institute of Technology, Cambridge, MA 02139, USA}
\affil[2]{Department of Materials Science and Engineering,
  Massachusetts Institute of Technology, Cambridge, MA 02139, USA}
\affil[3]{Department of Chemistry, Emory University, Atlanta, GA 30322, USA}
\affil[4]{Department of Nuclear Science and Engineering,
  Massachusetts Institute of Technology, Cambridge, MA 02139, USA}
\affil[5]{Honda Research Institute USA, San Jose, CA 95134, USA}

\date{}

\begin{document}

\twocolumn[
  \begin{@twocolumnfalse}
    \maketitle
    \begin{abstract}\noindent
Coupled-cluster theory defines the accuracy standard for molecular
electronic-structure properties but scales too steeply for routine application,
whereas density-functional theory is affordable yet systematically biased. We
resolve this trade-off with a single equivariant network, HARP (Hamiltonian Read-out for Properties), that
predicts an effective one-electron Hamiltonian from one inexpensive
B3LYP/def2-SVP calculation and derives a broad suite of properties from it
(energy, optical gap, dipole, quadrupole, polarizability, Mulliken
atomic charges, and Mayer bond orders) at coupled-cluster accuracy across nine main-group
elements, including the under-served phosphorus, sulfur, and chlorine
chemistries. The model is trained on a new in-house dataset of multi-property
labels computed at the CCSD(T) level for all nine elements. On a held-out test
set, it reduces the error of every property by a factor of $3.8$ to $270$
relative to semi-local, hybrid, and double-hybrid DFT (referenced to
composite CCSD(T)/cc-pVTZ; Methods),
while adding only $\sim$0.1~s wall time per molecule, delivering coupled-cluster-quality
predictions at the cost of a single DFT calculation. Critically, deriving every
property from a predicted Hamiltonian rather than pooling per-atom features
builds the correct size-scaling into the model architecture: on
$\pi$-conjugated oligothiophenes it matches finite-field CCSD polarizability
to $\sim$1\% and the EOM-CCSD optical gap to $\sim$3\% at the largest sizes where those
references remain affordable ($44$ and $37$ atoms, where a single CCSD field
point already costs $\sim$500$\times$ the model's entire inference) and
extrapolates the corrected trends to 58-atom chains, a regime where
pooling-based architectures fail by construction.
Accurate extrapolation is therefore set by the model's inductive bias rather
than by the training data.
    \end{abstract}
    \vspace{0.6em}
  \end{@twocolumnfalse}
]

{\renewcommand\thefootnote{\fnsymbol{footnote}}%
 \footnotetext[1]{These authors contributed equally.}%
 \footnotetext[2]{Correspondence: \texttt{haot@mit.edu}}%
 \footnotetext[3]{Correspondence: \texttt{liju@mit.edu}}}%
\setcounter{footnote}{0}

Predicting the electronic properties of molecules from their atomic structure is a
foundational capability of the chemical sciences, underpinning the rational
design of drugs, catalysts, electrolytes, and functional
materials~\cite{curtarolo2013high,butler2018machine}. As computational
screening increasingly precedes and guides experiments, the rate of discovery is
set by how accurately and inexpensively electronic-structure properties can be
evaluated across the vast space of candidate
compounds~\cite{vonlilienfeld2020exploring}. A method that combines the accuracy
of high-level wavefunction theory with a cost low enough for routine
application to large molecular libraries would transform this process. Developing such approaches has therefore long been a central
goal of computational chemistry.

The trade-off between accuracy and efficiency is the clearest between high-level wavefunction-based quantum chemistry method and density-functional theory. Coupled-cluster theory with perturbative triples, CCSD(T), is the \textit{de facto} gold standard for single-reference
molecules~\cite{raghavachari1989ccsdt}, but its $\mathcal{O}(N^7)$ scaling
confines it to small systems and precludes high-throughput use. Density-functional theory~\cite{kohn1965self} is far more affordable and is the most widely applied electronic-structure method, yet its accuracy is
functional-dependent and subject to systematic errors that no single functional
eliminates. Generalized-gradient approximations, in particular, exhibit a
delocalization (self-interaction) error that degrades response properties and
worsens with extended
$\pi$-conjugation~\cite{cohen2008insights,mori-sanchez2006localization}.
Ascending the hierarchy of functionals reduces some errors while introducing
others, at steadily increasing cost. Decades of methodological development have
narrowed, but not closed, the gap between affordable and accurate
electronic-structure methods.

Machine learning offers a route around this trade-off, because a trained
model's inference cost is set by its \emph{input}, not by its supervision: a
network trained on coupled-cluster labels whose input is a single
inexpensive DFT calculation (here, a $\Delta$-learning correction to the DFT
Fock matrix) reproduces correlated-level accuracy at DFT cost. Two broad
strategies have been pursued, and
each, we argue, addresses only one half of the problem. The first learns the
electronic structure itself, regressing the Kohn--Sham Hamiltonian from
geometry and then diagonalizing it, so that properties follow from quantum
mechanics with the correct dependence on system size. This route has been
developed for periodic solids by DeepH and its equivariant
successors~\cite{li2022deeph,gong2023deephe3}. For molecules, related approaches include
SchNOrb~\cite{schutt2019schnorb}, PhiSNet~\cite{unke2021phisnet},
QHNet~\cite{yu2023qhnet}, and, more recently, HELM, which spans much of the
periodic table~\cite{kaniselvan2025helm}. Because the supervision is the
reference Hamiltonian matrix, however, the accuracy is capped at the functional
that produced the labels, and the model accelerates DFT rather than surpassing
it.

The second strategy learns the properties directly. Equivariant property
networks such as SpookyNet~\cite{unke2021spookynet},
PaiNN~\cite{schutt2021painn}, and AIMNet2~\cite{anstine2024aimnet2}, and
$\Delta$-learning models more broadly~\cite{ramakrishnan2015delta}, can be
trained against correlated or experimental references and so are not capped in
accuracy. Their limitation lies instead in how each property is produced: it is
read out of pooled atomic features, a sum or mean over the graph. This
pooling fixes the size-scaling of every property to be strictly extensive or
strictly intensive, so a response that grows super-linearly with system size
cannot be represented, and accuracy degrades once the model is applied beyond
its training sizes. A separate output head must also be trained for each
property, in contrast to the single electronic-structure object from which an
entire suite of observables follows by construction~\cite{schutt2019schnorb,unke2021phisnet}. Most of these models are, moreover, confined to first-
and second-row organic chemistry (C, H, N, O). Indeed, Si, P, S, and Cl on the same
footing are rare, and many widely used potentials omit P and S entirely.

A third strategy improves the exchange--correlation functional itself:
deep-learning functionals such as DM21~\cite{kirkpatrick2021dm21} and
Skala~\cite{msr2025skala} attain
near-coupled-cluster accuracy for \emph{energies} at semi-local cost and span
the main group. Their supervision and validated benchmarks are, however,
energetic (DM21 additionally constrains fractional-charge and fractional-spin
behaviour): every other observable still follows from the resulting Kohn--Sham
solution, and accuracy at the correlated level for densities, response
properties, or excitations is neither trained for nor yet demonstrated.

Our model occupies the intersection not addressed by these strategies. Like
MEHnet, the Hamiltonian-learning approach introduced for hydrocarbons by
Tang \emph{et al.}~\cite{tang2024multitask}, our model---HARP, the Hamiltonian Read-out for Properties---predicts an effective one-electron Hamiltonian and derives every
property from it using exact quantum-mechanical operators, inheriting the correct
size-scaling by construction. Critically, it is supervised on the
\emph{observables} rather than on a reference Hamiltonian matrix: a single
learned correction to one cheap B3LYP/def2-SVP single point is trained so that
the derived properties match a correlated reference; the model is therefore not
capped at DFT accuracy and could, in principle, target any level of theory. Our contributions are fourfold:
\textbf{(i)} deriving \emph{every} property from the predicted Hamiltonian by
exact operators builds the correct size-scaling into the architecture itself,
and the model demonstrably extrapolates out of distribution: it tracks
finite-field CCSD polarizability on $\pi$-conjugated oligothiophenes to
$\sim$1\% out to the 44-atom CCSD limit and continues the corrected trend to
58-atom chains, all from sub-25-atom training molecules, where pooling-based
architectures diverge (Fig.~\ref{fig:scaling}), because the inductive bias,
not the data, sets the size-scaling;
\textbf{(ii)} it reaches CCSD(T)/CCSD/EOM-CCSD-level accuracy on seven ground-
and response-properties at the cost of one DFT single point, all from one
baseline rather than a per-property construction;
\textbf{(iii)} it spans nine
main-group elements, treating the under-served P/S/Cl chemistries on the same
footing as the organic elements; and
\textbf{(iv)} it does so with an architecture built for the main group: a
higher-angular-resolution equivariant backbone (EquiformerV2, $l_{\max}=4$, as
required by the $d$-orbital Fock blocks of the third-row elements),
$\Delta$-learning extended to \emph{every} predicted object---an explicit
energy-correction head and a screening matrix anchored to the baseline
response, so that an untrained correction reproduces the baseline
polarizability exactly---and a hybrid baseline whose near-unit-slope
Kohn--Sham gap lets the optical gap be read directly from the frontier
eigenvalues, with no learned rescaling (SI).

Two principles organize these results: the model's architecture determines
whether a property extrapolates beyond the training distribution, and the level
of supervision sets the accuracy it can reach. Together they place a broad suite
of molecular properties at the favourable corner of the accuracy--cost plane
across the nine main-group elements covered here.

\section{Results}
\label{sec:results}

\subsection{Model architecture and dataset}

\begin{figure}[!tb]\centering
  \includegraphics[width=\linewidth]{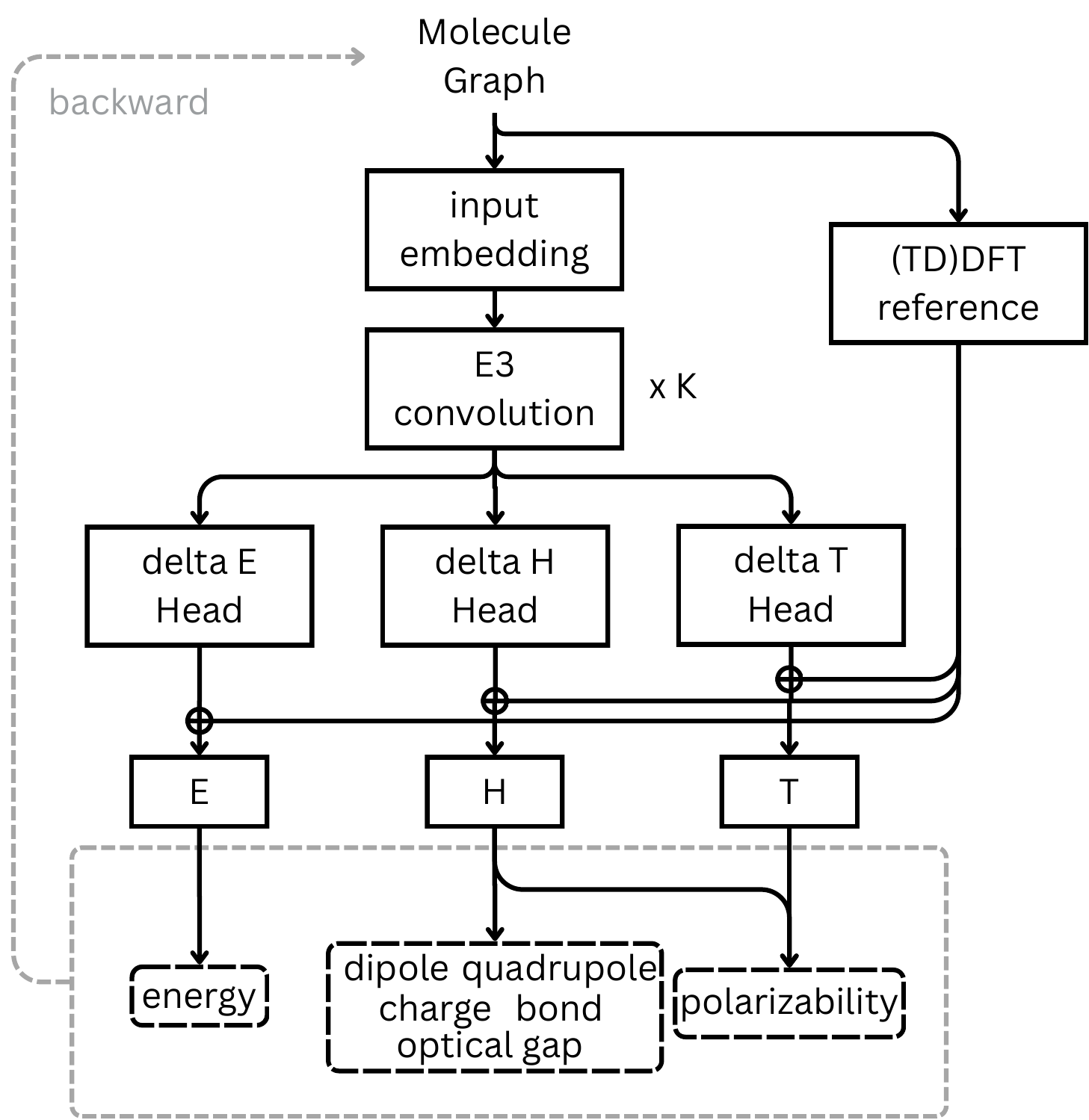}
  \caption{\textbf{Model overview.} From a molecule graph, an
  equivariant network (input embedding followed by $K$ E(3)-convolution layers)
  produces node and edge features that feed three correction heads:
  $\Delta E_\theta$, $\Delta\mathbf{H}_\theta$, and $\Delta\mathbf{T}_\theta$.
  Each correction is added to the corresponding baseline ($E_{\mathrm{DFT}}$, $\mathbf{H}_{\mathrm{DFT}}$,
  $\mathbf{T}_{\mathrm{DFT}}$) from the cheap (TD)DFT (B3LYP/def2-SVP) reference to give the
  total energy $E$, the effective Hamiltonian $\mathbf{H}$, and the screening
  matrix $\mathbf{T}$. Properties are then read off
  by the defining quantum-mechanical operator rather than pooled from atomic
  features: the energy from $E$; the dipole, quadrupole, Mulliken atomic
  charges, Mayer bond
  orders, and optical gap from $\mathbf{H}$; and the polarizability from the
  screening matrix $\mathbf{T}$. The dashed path denotes gradients (response properties obtained
  by differentiation). The full neural inference adds only $\sim$0.1~s (GPU) on top
  of the DFT baseline, so coupled-cluster-quality properties are obtained at the
  cost of one DFT single point.}
  \label{fig:overview}
\end{figure}
\label{sec:coverage}

Rather than correcting each target property directly, our model applies a
learned correction to a cheap physical baseline at the level of a few underlying
electronic-structure objects, and derives every property from them
(Fig.~\ref{fig:overview}). A single B3LYP/def2-SVP
calculation supplies the baseline: a total energy $E_{\mathrm{DFT}}$, a Fock matrix
$\mathbf{H}_{\mathrm{DFT}}$, and a screening matrix $\mathbf{T}_{\mathrm{DFT}}$. The
ground-state single point provides $E_{\mathrm{DFT}}$ and
$\mathbf{H}_{\mathrm{DFT}}$, and hence every property except the
polarizability; only $\mathbf{T}_{\mathrm{DFT}}$ requires an additional
linear-response (TDDFT) step on the same baseline, so the TDDFT cost is
incurred only when $\boldsymbol{\alpha}$ is requested. An
E(3)-equivariant graph network reads the molecule graph and produces, through
separate heads, three corrections to this baseline, $\Delta E_\theta$,
$\Delta\mathbf{H}_\theta$, and $\Delta\mathbf{T}_\theta$, which are added back to
give a corrected energy $E$, an effective Hamiltonian $\mathbf{H}$, and a
screening matrix $\mathbf{T}$. Its backbone is a spherical-harmonic
graph-attention network (EquiformerV2~\cite{liao2024equiformerv2};
$\sim$5.6~M parameters, four equivariant interaction layers,
spherical-harmonic degree $l_{\max}=4$, a $6$~\AA{}
cutoff); on top of this backbone we add equivariant heads, built with
e3nn~\cite{geiger2022e3nn}, that output the three
corrections. Full settings are given in Methods. The
essential design choice is that the network outputs corrections to
these physical objects, not the target properties themselves. Because the
baseline is inexpensive and the network adds only milliseconds
(Fig.~\ref{fig:bulk}a), the total cost is essentially that of the hybrid-DFT
single point, while the accuracy approaches the coupled-cluster labels the
network is trained on.

A single such model spans all nine main-group elements (H, C, N, O, F, Si, P,
S, Cl) and yields seven properties from one forward pass: the total electronic
energy $E$, the gap $E_g$, and Mulliken atomic
charges~\cite{mulliken1955population} $q_i$ (scalars); the dipole
$\boldsymbol{\mu}$ (a vector); the quadrupole $\mathbf{Q}$ and polarizability
$\boldsymbol{\alpha}$ (rank-2 tensors, predicted equivariantly); and pairwise
Mayer bond orders~\cite{mayer1983charge} $b_{ij}$. Each corrected object yields its properties by the
operator that defines it, rather than through a separate property head. The corrected energy
$E=E_{\mathrm{DFT}}+\Delta E_\theta$ is the total electronic energy. Diagonalizing
the effective Hamiltonian
$\mathbf{H}=\mathbf{H}_{\mathrm{DFT}}+\Delta\mathbf{H}_\theta$ gives molecular-orbital
energies and coefficients, and hence the one-particle density matrix, from
which follow the optical gap (the frontier HOMO--LUMO eigenvalue difference),
the dipole and quadrupole (multipole operators traced against the density), and
the atomic charges and bond orders. The polarizability uses
the corrected screening matrix
$\mathbf{T}=\mathbf{T}_{\mathrm{DFT}}+\Delta\mathbf{T}_\theta$: a bare
sum-over-states response built from the eigenstates of $\mathbf{H}$ is screened
by $\mathbf{T}$ to give the reported tensor (Methods,
Eqs.~\eqref{eq:sos}--\eqref{eq:screen}), and because $\mathbf{T}_{\mathrm{DFT}}$ comes
from the baseline DFT response, an uncorrected model reproduces the baseline
polarizability. Because each property is the exact functional of objects
assembled from \emph{local} corrections, its dependence on system size is fixed
by quantum mechanics rather than by a pooling rule, which is what allows the
model to extrapolate collective responses beyond the training sizes
(Fig.~\ref{fig:scaling}).

HARP keeps this effective-Hamiltonian read-out from its hydrocarbon
predecessor~\cite{tang2024multitask} but rebuilds the machinery around it.
The two-layer EGNN backbone of MEHnet ($l_{\max}=2$) becomes the four-layer
EquiformerV2 above, carrying angular momenta to $l_{\max}=4$ as the $d\times d$
Fock blocks of Si--Cl require; the correction set gains an explicit energy head
$\Delta E_\theta$ (MEHnet read the energy as the occupied-eigenvalue sum); the
screening matrix becomes a $\Delta$-learning quantity anchored to the baseline
response, $\mathbf{T}=\mathbf{T}_{\mathrm{DFT}}+\Delta\mathbf{T}_\theta$,
where MEHnet learned $\mathbf{T}$ from scratch; and the hybrid baseline allows the
gap to be the bare frontier eigenvalue difference, retiring MEHnet's learned gap
rescaling, $E_g=(1+G_1)(\varepsilon_{\mathrm{L}}-\varepsilon_{\mathrm{H}})+G_2$.
Retiring the rescaling is more than a simplification. The $G_1,G_2$ correctors
were attention-pooled learned scalars---precisely the pooling-style read-out
this paper argues has no guaranteed size behaviour---and they were the last
property that bypassed the Hamiltonian. With them removed, every property in
HARP flows through the operator read-out, so the size transferability above
applies to the full suite; the out-of-distribution gap of
Fig.~\ref{fig:scaling}b is its direct test. A row-by-row comparison with MEHnet,
including accuracy, is given in the SI.

\begin{figure*}[!tb]\centering
  \includegraphics[width=\textwidth]{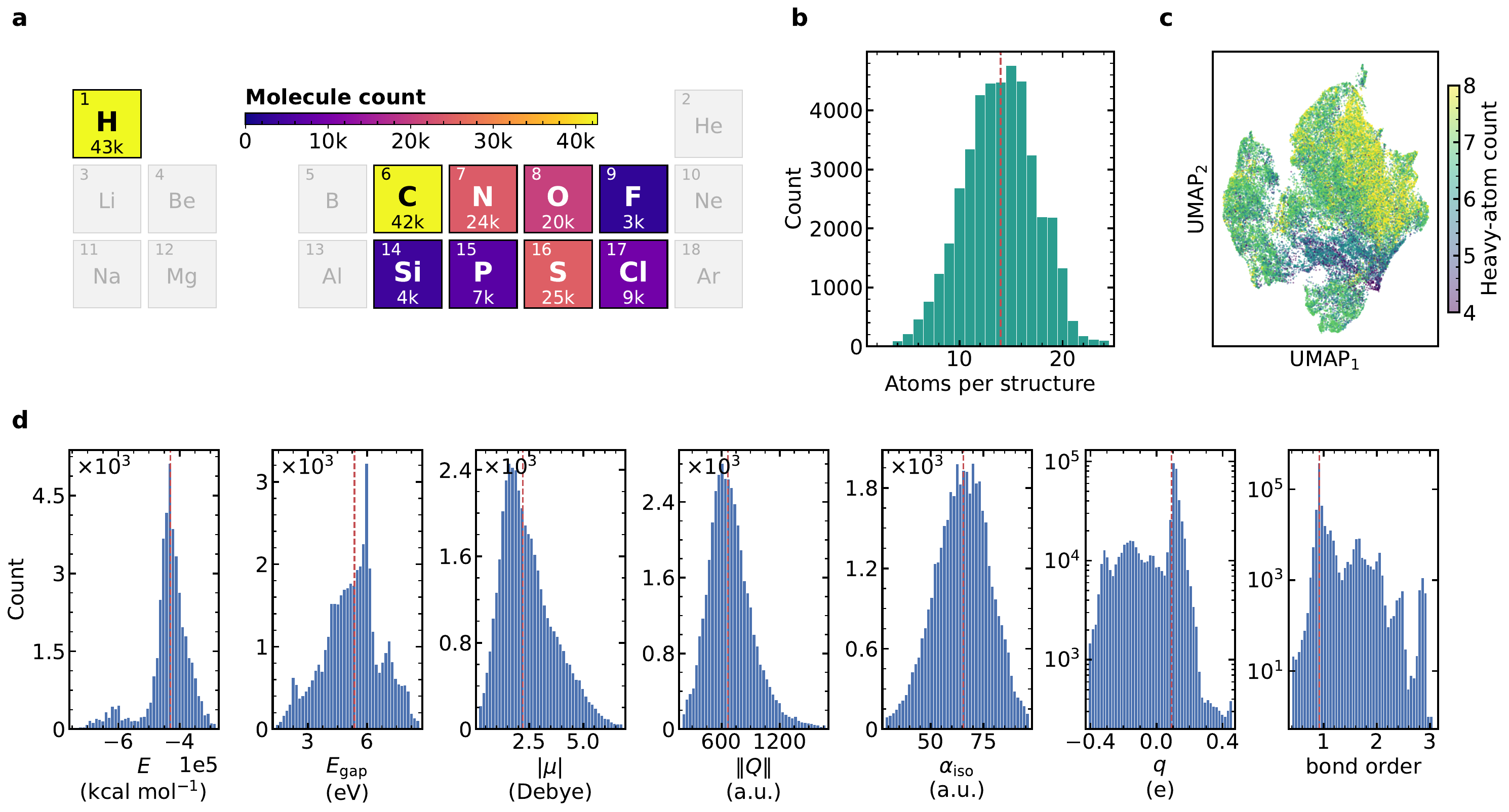}
  \caption{\textbf{Dataset composition and coverage.} \textbf{a} Elemental composition of the dataset across the nine main-group elements included in this work (periods 1--3). Shaded as a heatmap to show the number of molecules in the dataset containing a given element. \textbf{b} Distribution of molecular size, measured as atoms per structure, with median size $14$. \textbf{c} UMAP embedding of the CCSD(T)
  property space, colored by heavy-atom count. \textbf{d} Distributions of the seven CCSD(T) ground-truth properties across the dataset: total ground state energy $E$, the optical gap $E_\mathrm{gap}$, the dipole $|\boldsymbol{\mu}|$, the
  quadrupole $\|\mathbf{Q}\|$, the isotropic polarizability
  $\alpha_\mathrm{iso}$, the per-atom Mulliken charge $q$, and the per-bond Mayer bond order (dashed lines mark medians).}
  \label{fig:dataset}
\end{figure*}

The model is trained on a new in-house dataset built from $44{,}412$ three-dimensional structures from PubChem and passed through a quality filter (Methods). Designed to broaden chemical coverage, the dataset includes compounds containing silicone, phosphorus, sulfur, chlorine, and fluorine in addition to the organic elements H, C, N, O. (Fig.~\ref{fig:dataset}a). The retained molecules are small, neutral, closed-shell singlets of $2$--$24$ atoms ($\le 9$ heavy, median $14$; Fig.~\ref{fig:dataset}b). In terms of element coverage, hydrogen ($99.3\%$ of molecules) and carbon ($98.1\%$) are near-ubiquitous. The set is also rich in the heteroatoms that most omit: sulfur in $58.1\%$ of molecules, nitrogen in $56.8\%$, oxygen in $47.6\%$, chlorine in $21.9\%$, phosphorus in $15.2\%$, silicon in $9.5\%$, and fluorine in $6.5\%$ (Fig.~\ref{fig:dataset}a). Every molecule is labeled at
property-appropriate coupled-cluster levels (energies at composite
CCSD(T)/cc-pVTZ---canonical CCSD(T)/cc-pVDZ plus a DLPNO-CCSD(T) basis-set
correction to cc-pVTZ, Eq.~\eqref{eq:composite}---polarizabilities at
finite-field CCSD/cc-pVDZ, gaps at EOM-CCSD/cc-pVDZ, and dipoles, quadrupoles,
Mulliken charges and Mayer bond orders from the coupled-cluster
density at the same composite level; Methods), and the resulting ground-truth properties span a broad chemical range (Fig.~\ref{fig:dataset}c,d). After filtering, $42{,}912$ molecules remain. These are partitioned into training, validation, and a held-out test set of $959$ molecules, which we report on throughout this work.

\subsection{Accuracy benchmark across all properties}

\begin{table*}[!tb]\centering
  \caption{\textbf{Global mean absolute error (MAE) versus the
  composite CCSD(T)/cc-pVTZ ground truth (Methods)} on the 959-molecule test set, per
  property and method. Lower is better; best in bold. Units: $E$ in
  kcal\,mol$^{-1}$, $E_\mathrm{gap}$ in eV, $|\mu|$ in Debye; $\|Q\|$ and
  $\alpha_\mathrm{iso}$ in a.u.; charge in $e$. Each method is evaluated at the
  basis set of the property's coupled-cluster reference: cc-pVTZ for $E$
  (compared as an atomization energy), $|\mu|$, $\|Q\|$, $q$ and bond order, and
  cc-pVDZ for $\alpha_\mathrm{iso}$ and $E_\mathrm{gap}$ (finite-field CCSD and
  EOM-CCSD, respectively). DSD-PBEP86 is the D3(BJ)-corrected double hybrid.}
  \label{tab:mae}
  \vspace{1em}
  \small
  \begin{tabular}{lcccc}
    \toprule
    property & BP86 & B3LYP & DSD-PBEP86 & \textbf{HARP (this work)} \\
    \midrule
    $E$ (kcal\,mol$^{-1}$) & 63.0 & 13.2 & 15.1 & \textbf{0.23} \\
    $E_\mathrm{gap}$ (eV) & 1.32  & 0.61  & 4.84  & \textbf{0.067} \\
    $|\mu|$ (Debye)       & 0.160 & 0.159 & 0.109 & \textbf{0.022} \\
    $\|Q\|$ (a.u.)        & 0.389 & 0.347 & 0.233 & \textbf{0.044} \\
    $\alpha_\mathrm{iso}$ (a.u.) & 4.71 & 2.79 & 1.18 & \textbf{0.121} \\
    $q$ (e)               & 0.0205 & 0.0256 & 0.0220 & \textbf{0.0053} \\
    bond order            & 0.0556 & 0.0588 & 0.0251 & \textbf{0.0066} \\
    \bottomrule
  \end{tabular}
\end{table*}

\begin{figure*}[!tb]\centering
  \includegraphics[width=0.88\textwidth]{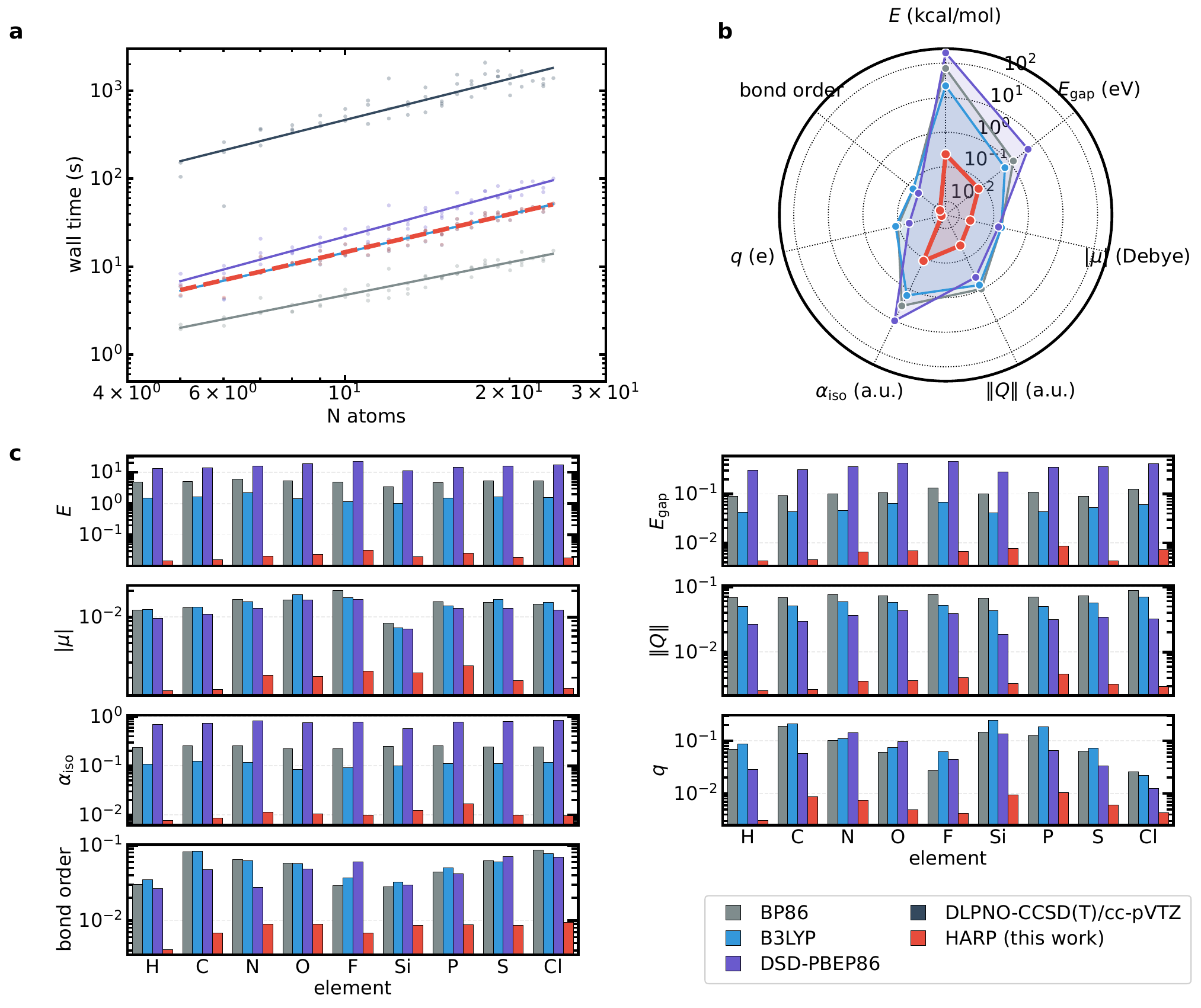}
  \caption{\textbf{Bulk accuracy and cost.} \textbf{a} Per-molecule wall time
  per single-point evaluation versus atom count (log--log) with power-law fits;
  the model shares its DFT baseline's shallow scaling, while CCSD(T) and the
  double-hybrid scale steeply. \textbf{b} Per-property error radar on a shared
  log scale (inner $\to$ smaller error vs.\ CCSD(T)). \textbf{c} Per-element MAE
  across the nine supported elements, per-atom averaged; units: Mulliken charge in $e$,
  $E$ in kcal\,mol$^{-1}$/atom, $E_\mathrm{gap}$ in eV, $\mu$ in D/atom, $Q$ and
  $\alpha_\mathrm{iso}$ in a.u./atom. As in Table~\ref{tab:mae}, each functional
  is evaluated at the basis of the property's coupled-cluster reference
  (cc-pVTZ for $E$, $|\mu|$, $\|Q\|$, $q$ and bond order; cc-pVDZ for
  $\alpha_\mathrm{iso}$ and $E_\mathrm{gap}$), all referenced to
  composite CCSD(T)/cc-pVTZ (Methods), finite-field CCSD/cc-pVDZ and
  EOM-CCSD/cc-pVDZ; panel
  \textbf{a} times the cc-pVTZ single point of each functional against
  DLPNO-CCSD(T)/cc-pVTZ, the cc-pVTZ component of the composite reference.}
  \label{fig:bulk}
\end{figure*}
\label{sec:bulk}

On the 959-molecule test set, we compare the model against three widely used
density functionals (GGA BP86, hybrid B3LYP, and double-hybrid DSD-PBEP86). Each functional is evaluated at the basis set of the property's coupled-cluster reference (cc-pVTZ for energies, dipoles, quadrupoles, Mulliken charges and Mayer bond orders; cc-pVDZ for polarizabilities and optical gaps), referenced to the composite CCSD(T)/cc-pVTZ, finite-field CCSD/cc-pVDZ, and EOM-CCSD/cc-pVDZ ground truth
(Table~\ref{tab:mae}, Fig.~\ref{fig:bulk}). The model reduces the mean absolute error of
\emph{every} property relative to the DFT methods by factors ranging from
$\sim$$2.5$ (per-atom charges and bond orders) to over two orders of
magnitude (energies). The total energy reaches sub-chemical accuracy
($\sim$$10^{-3}$~Ha, well below $1$~kcal\,mol$^{-1}$), more than two orders of
magnitude below BP86, and the tensorial and response properties (quadrupole,
polarizability) improve by roughly an order of magnitude, the regime where DFT
performs worst. Simply ascending the functional hierarchy does not close this
gap, whereas the learned correction does: on this set B3LYP is not consistently
better than BP86, while the double-hybrid DSD-PBEP86 --- each functional now
evaluated at a basis matched to its coupled-cluster reference --- is the most
accurate of the three for the tensorial and response properties (dipole,
quadrupole, polarizability, and bond order), consistent with its design, whereas
B3LYP is best for the energy. The double hybrid's Kohn--Sham orbital gap,
however, remains a poor proxy for the EOM-CCSD optical gap. HARP
nonetheless leads every property in the table. Deriving every property from one predicted
Hamiltonian by exact operators is what lets HARP report the complete suite
at one accuracy level. The correction also generalizes uniformly
across chemistry, with no degradation for the rarer elements F and Si relative
to C/H/N/O (Fig.~\ref{fig:bulk}c), evidence that it has learned transferable
chemistry rather than memorizing the common organic motifs. The hydrocarbon-only
predecessor MEHnet~\cite{tang2024multitask} is compared per-property in the
Supplementary Information (Tables~S11 and~S12); because MEHnet reports in-distribution
RMSE on hydrocarbons against a cc-pVDZ reference, it is not directly
commensurable with the columns here and is kept out of Table~\ref{tab:mae}.

This accuracy comes at the cost of using the B3LYP baseline.
Figure~\ref{fig:bulk}a reports the per-molecule wall time of a single-point
energy at each method's reference level (serial ORCA). Fitted
as $t = A\,N^{\tau}$, the DLPNO-CCSD(T)/cc-pVTZ component of the composite
reference is one to two orders
of magnitude more expensive than the DFT baselines at every size in the set
(prefactor $A\approx13$~s versus $0.3$--$0.5$~s) and is the steepest-scaling method
shown ($\tau\approx1.6$), reaching
$\sim$$10^{3}$~s on the largest molecules. These fitted exponents are
\emph{effective} slopes over the $5$--$24$-atom test range and sit well below
the familiar formal asymptotics ($\mathcal{O}(N^{3\text{--}4})$ for hybrid DFT,
$\mathcal{O}(N^{7})$ for canonical CCSD(T)): at these sizes the wall time is
dominated by size-independent overheads (SCF setup, integral generation, I/O)
and reduced by integral screening and density fitting (RIJCOSX), and
DLPNO-CCSD(T) in particular trades the canonical $\mathcal{O}(N^{7})$ for
near-linear asymptotic scaling at a $\sim$$30\times$ prefactor, appearing as a
vertically offset, nearly parallel line rather than a steeper curve. The
textbook exponents only emerge at larger sizes: on the $9$--$44$-atom
oligothiophenes the measured canonical-CCSD slope steepens to
$\tau\approx4.6$ (Supplementary Fig.~S5). The model rides its B3LYP/def2-SVP
baseline: the neural inference (network forward pass, diagonalization, and
property evaluation) adds only $0.10$--$0.14$~s per molecule on a single
NVIDIA A100 GPU, nearly flat in size, so its cost tracks the baseline DFT
curve, roughly two orders below CCSD(T); the deep-learning functional
Skala~\cite{msr2025skala} is comparably cheap (a meta-GGA-cost single point). We
report wall-times because they are what a practitioner experiences, but they
conflate hardware (ORCA on CPU, the network and Skala on GPU); expressed as
accelerator-time the model's cost is dominated by the classical baseline either
way (SI). The model therefore sits at the favourable corner of the
accuracy--cost plane (Fig.~\ref{fig:overview}d): coupled-cluster accuracy at
hybrid-DFT cost, $\sim$two orders below CCSD(T) at these sizes and a separation
that widens with system size, without bound once the reference turns
intractable, as in the larger oligothiophenes below.

Accuracy aside, no released machine-learning model even delivers this
combination of chemistry and properties. Table~\ref{tab:mlcompare} maps
representative released models onto the benchmark's nine-element chemistry and
property suite: across the three families of ML approaches to molecular
properties, none simultaneously spans the elements, outputs the full property
suite, and targets a correlated reference. A complementary scope map---which
charge states, spin states, and geometry regimes each of these models is
trained or validated on---is given in the Supplementary Information.

\providecommand{\tyes}{\textcolor{black}{\checkmark}}
\providecommand{\tcon}{\textcolor{black!42}{\checkmark}}
\providecommand{\tno}{\textcolor{black!22}{\textendash}}
\providecommand{\trel}{\textcolor{black}{\checkmark}}
\providecommand{\tnorel}{\textcolor{black!55}{\ensuremath{\times}}}
\begin{table*}[!tb]\centering
  \caption{\textbf{What released machine-learning models can deliver on this
  benchmark's chemistry and property suite.} A capability/coverage map, not a
  head-to-head error table (for accuracy see Table~\ref{tab:mae}): across the
  three families of ML approaches to molecular properties, no released model
  simultaneously spans the nine-element chemistry, outputs the full property
  suite, and targets a correlated reference. \emph{Elements} states each
  released model's own coverage---several exceed this work's scope---with, in
  parentheses, how many of this benchmark's nine elements (H, C, N, O, F, Si,
  P, S, Cl) it spans: QHNet lacks Si, P, S, Cl; the released SpookyNet
  (QM7-X) lacks F, Si, P; MACE-OFF23 lacks Si.
  Symbols: \tyes~validated output of the cited model;
  \tcon~obtainable from the model's predicted electronic-structure objects
  (Hamiltonian, density, or charges) but not demonstrated in the cited work;
  \tno~not accessible. Properties: total/atomization energy $E$, HOMO--LUMO gap
  $E_\mathrm{g}$, dipole $\mu$, quadrupole $Q$, polarizability $\alpha$, atomic
  charge $q$, bond order (BO); this work reports Mulliken charges and Mayer bond
  orders, whereas the population convention differs among the other entries
  (SpookyNet, e.g., predicts its own learned partial charges).
  HELM's weights are announced but, as of this
  writing, not public (its dataset is); MACE-OFF23 dipoles are validated only
  in its separate $-\mu$ variant; SpookyNet dipoles follow from its predicted
  partial charges but are not benchmarked.
  The \emph{Reference} column is the correlated level the tabulated properties are
  fit to: for the machine-learned functionals (DM21, Skala) only the energy
  targets a correlated reference ($E$), whereas HARP targets one for
  \emph{every} property (all).}
  \label{tab:mlcompare}
  \vspace{1em}
  \small
  \setlength{\tabcolsep}{5pt}
  \renewcommand{\arraystretch}{1.32}
  \begin{tabular}{l c *{7}{c} c c}
    \toprule
    & & \multicolumn{7}{c}{Output properties} & & \\
    \cmidrule(lr){3-9}
    Method & Elements & $E$ & $E_\mathrm{g}$ & $\mu$ & $Q$ & $\alpha$ & $q$ & BO & Reference & Public \\
    \midrule
    \multicolumn{11}{l}{\textit{Electronic-structure (Hamiltonian) read-out}}\\
    \rowcolor{black!7}
    \textbf{HARP (this work)} & \textbf{9 (9/9)}
      & \tyes & \tyes & \tyes & \tyes & \tyes & \tyes & \tyes
      & \textbf{CCSD(T) (all)} & \trel \\
    QHNet~\cite{yu2023qhnet}       & 5 (5/9)
      & \tcon & \tcon & \tcon & \tcon & \tcon & \tcon & \tcon
      & B3LYP & \trel \\
    HELM~\cite{kaniselvan2025helm} & 58 (9/9)
      & \tyes & \tcon & \tcon & \tcon & \tcon & \tcon & \tcon
      & $\omega$B97M-V & \tnorel \\
    \addlinespace[2pt]
    \multicolumn{11}{l}{\textit{Machine-learned density functional}}\\
    DM21~\cite{kirkpatrick2021dm21} & H--Kr (9/9)
      & \tyes & \tcon & \tcon & \tcon & \tcon & \tcon & \tcon
      & CCSD(T)/exp.\ ($E$) & \trel \\
    Skala~\cite{msr2025skala}      & H--Ar (9/9)
      & \tyes & \tcon & \tyes & \tcon & \tcon & \tcon & \tcon
      & CCSD(T) ($E$) & \trel \\
    \addlinespace[2pt]
    \multicolumn{11}{l}{\textit{Direct property (pooling) read-out}}\\
    SpookyNet~\cite{unke2021spookynet}  & 6 (6/9)
      & \tyes & \tno & \tcon & \tno & \tno & \tyes & \tno
      & PBE0+MBD & \trel \\
    MACE-OFF23~\cite{kovacs2025maceoff} & 10 (8/9)
      & \tyes & \tno & \tcon & \tno & \tno & \tno & \tno
      & $\omega$B97M-D3(BJ) & \trel \\
    UMA / OMol25~\cite{wood2025uma}     & 83 (9/9)
      & \tyes & \tno & \tno & \tno & \tno & \tno & \tno
      & $\omega$B97M-V & \trel \\
    \bottomrule
  \end{tabular}
\end{table*}

\subsection{Agreement with experiment}
\label{sec:dipole}

The benchmarks so far are referenced to a computational ground truth. As an
independent check, we compare predicted gas-phase dipole magnitudes against
\emph{experimentally measured} values from the NIST Computational Chemistry Comparison and
Benchmark Database (CCCBDB), for twelve molecules spanning the full element
scope (HF, HCl, H$_2$S, PH$_3$, SiH$_4$, CH$_3$F, CH$_3$Cl, CHF$_3$,
CHCl$_3$, PF$_3$, PCl$_3$, SO$_2$; experimental dipoles $0$--$1.89$~Debye).
We use the dipole moment because it is sensitive to electronic-structure
quality yet insensitive to basis-set incompleteness in the cc-pV$X$Z series,
so the experimental value is a clean target
(unlike atomization energies, where cc-pVTZ incompleteness dominates;
discussed in the SI).

\begin{figure}[!tb]\centering
  \includegraphics[width=\linewidth]{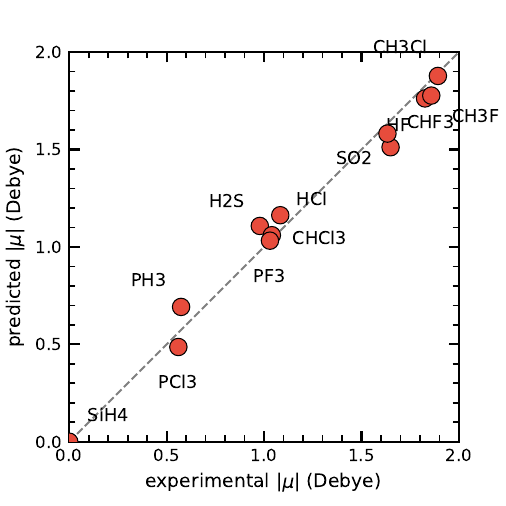}
  \caption{\textbf{Agreement with experiment.} Predicted versus measured
  (NIST CCCBDB) gas-phase dipole magnitudes for twelve molecules spanning
  the model's element scope; the line is $y=x$. Mean absolute error
  $0.064$~Debye (RMSE $0.079$~Debye), compared with the B3LYP baseline
  and $0.046$~Debye for a canonical CCSD(T)/def2-TZVPP reference. SiH$_4$ is at the origin
  by $T_d$ symmetry.}
  \label{fig:dipole}
\end{figure}

Figure~\ref{fig:dipole} shows the parity plot against experiment. Over the
twelve molecules the model predicts the measured dipole magnitudes to a mean
absolute error of $0.064$~Debye---close to the $0.046$~Debye that a canonical
CCSD(T)/def2-TZVPP calculation achieves against the same experiments, and an
improvement over its B3LYP/def2-SVP baseline, at the baseline's cost. The model thus reproduces a real,
independently measured observable across H, C, N, O, F, Si, P, S, and Cl
chemistries at DFT cost.

\subsection{Out-of-distribution size scaling and its mechanism}

\begin{figure*}[!tb]\centering
  \includegraphics[width=\linewidth]{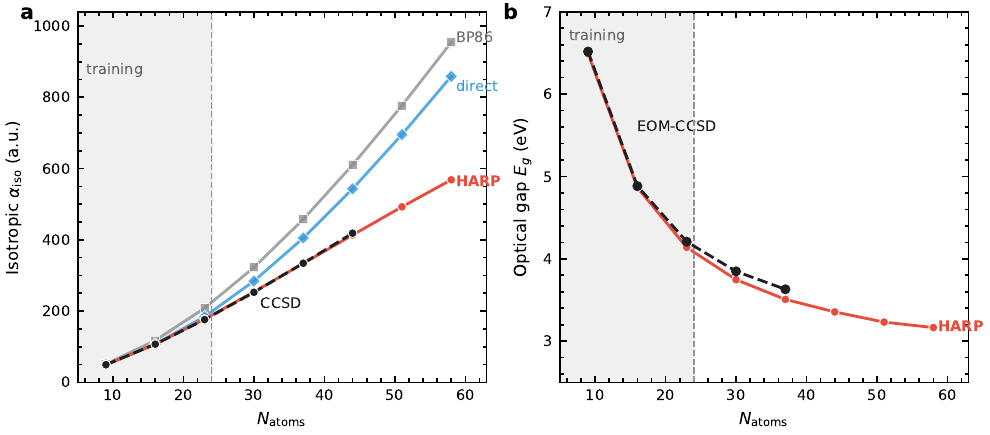}
  \caption{\textbf{Out-of-distribution size scaling on oligothiophenes
  $\mathrm{T}_n$ ($n=1$--$8$, $9$--$58$ atoms).} The direct read-out ablation
  (blue) shares the backbone architecture and dataset of HARP
  but predicts each property directly, as a correction to the BP86 baseline;
  a Hamiltonian read-out on the same BP86 baseline does not drift (SI). \textbf{a} Isotropic polarizability
  $\alpha_\mathrm{iso}$ versus size: the Hamiltonian read-out of this work
  (HARP, red) agrees with finite-field CCSD/cc-pVDZ (black, dashed) to
  within $1.3\%$ everywhere CCSD is computable ($n\le6$, up to 44 atoms) and
  continues the trend to 58 atoms, whereas the direct
  property read-out (blue; the pooling-style $\Delta$-prediction ablation)
  over-polarizes, drifting toward the BP86 baseline (grey). \textbf{b} Optical
  gap versus size: HARP (red) reproduces the EOM-CCSD/cc-pVDZ gap where that
  reference is affordable ($\mathrm{T}_1$--$\mathrm{T}_5$, black; within
  $0.03$~eV through $\mathrm{T}_2$ and within $0.13$~eV, $3.4\%$, at
  $\mathrm{T}_5$) and saturates with length; a Kohn--Sham (DFT) gap is not an optical gap and is not shown. The
  mechanism (frontier-orbital delocalization) and the longitudinal
  $\alpha_{xx}$ are analysed in the Supplementary Information.}
  \label{fig:scaling}
\end{figure*}
\label{sec:scaling}
\label{sec:scaling-theory}

The benchmarks above probe molecules from the training distribution. A more
demanding test is transfer to systems larger than every training molecule and
outside the training size distribution: $\alpha,\alpha'$-linked
oligothiophenes $\mathrm{T}_n$
($n=1$--$8$; thiophene to octithiophene, $9$--$58$ atoms), the sulfur-rich
$\pi$-conjugated motif of organic electronics. These are doubly
out-of-distribution: larger than the training molecules ($\le 9$ heavy and $24$
total atoms, already exceeded by bithiophene $\mathrm{T}_2$) and realising
extended $\pi$-conjugation absent from training, with coupled-cluster references
that turn intractable as the chain grows. The isotropic polarizability grows
super-linearly along the backbone (Fig.~\ref{fig:scaling}a; a $\sim$12-fold rise
from $\mathrm{T}_1$ to $\mathrm{T}_8$, $\sim$19-fold for the longitudinal
$\alpha_{xx}$, SI). HARP tracks finite-field CCSD/cc-pVDZ across the
affordable range (signed $\alpha_\mathrm{iso}$ error within $1.3\%$ out to
$\mathrm{T}_6$; the longitudinal $\alpha_{xx}$ stays within $2.3\%$, SI) and
continues the corrected trend to $\mathrm{T}_8$, while the
uncorrected DFT baseline over-polarizes by a length-growing margin ($+43\%$ at
$\mathrm{T}_4$ for a GGA). The decisive control is the \emph{read-out ablation}
(blue): a model with the same backbone architecture and dataset but a
direct property read-out (a per-property correction to the BP86 baseline, at
$l_{\max}=3$) reproduces CCSD on short chains but then drifts back toward the
BP86 baseline, the deviation widening monotonically to $\mathrm{T}_8$. The
baseline is not the cause: a Hamiltonian-read-out model on the same BP86
baseline stays within $4.3\%$ of CCSD in $\alpha_\mathrm{iso}$ through
$\mathrm{T}_5$, where the direct read-out is off by $21\%$ (SI). The
divergence is therefore attributable to the read-out.

The optical gap is the complementary observable (Fig.~\ref{fig:scaling}b): where
the polarizability grows super-linearly, the gap closes monotonically with
conjugation and saturates toward a finite polymer-limit value, an
intensive-like quantity that the same read-out also gets right. Our accuracy claim
here is deliberately narrow. The gap label is, by construction, the lowest
EOM-CCSD/cc-pVDZ excitation (Methods), so the like-for-like reference is
EOM-CCSD/cc-pVDZ on the same geometries; against it the model transfers well out
of distribution, reproducing the EOM-CCSD gap within $0.03$~eV at
$\mathrm{T}_1$--$\mathrm{T}_2$ and tracking the saturating trend thereafter
with a mild, size-growing underestimate that reaches $-0.12$~eV ($-3.4\%$) at
$\mathrm{T}_5$, the largest size where the reference remains affordable. Both the model and
this reference lie $\sim$1~eV above experimental UV--Vis maxima. That offset
originates in the limitations of the EOM-CCSD/cc-pVDZ labels, not in the
network: cc-pVDZ lacks diffuse functions, which alone account for $0.41$~eV of
the offset at $\mathrm{T}_1$ (SI), and the comparison further pits vertical
gas-phase excitations against solution band maxima. The model's gap accuracy is
therefore bounded by the level of its labels, a point we return to in the
Discussion.

A network with a finite cutoff $R_c$ can nevertheless capture a response that
grows over lengths far exceeding $R_c$, because each property is read from the
\emph{eigenstates} of the predicted Hamiltonian, not from local features:
although the learned correction $\Delta\mathbf{H}_\theta$ is local, diagonalizing
$\mathbf{H}=\mathbf{H}_{\mathrm{DFT}}+\Delta\mathbf{H}_\theta$ yields frontier
orbitals that delocalize over the conjugation length $\xi\gg R_c$, and the
polarizability is the linear response of these delocalized states (Methods,
Eqs.~\eqref{eq:sos}--\eqref{eq:screen}). This is measured, not assumed: the
frontier quantities of the predicted Hamiltonian keep evolving with chain
length well beyond the $6$~\AA{} radial cutoff and past the $24$~\AA{}
receptive field of the four-layer network, the HOMO$\to$LUMO transition dipole
growing as $\sim N^{0.68}$ and the gap shrinking as $\sim N^{-0.38}$ out to
$\mathrm{T}_8$ (SI); their
combination, $\alpha_{xx}\sim
|\langle\psi_H|\hat{x}|\psi_L\rangle|^2/(\varepsilon_L-\varepsilon_H)\sim N^{1.75}$,
is close to the $\sim N^{1.6}$ growth of the full predicted $\alpha_{xx}$:
super-linear ($\tau>1$). The response
length is set by the global eigenproblem ($\xi$), not by the network cutoff; the
full derivation is in the SI, and a distributed-polarizability decomposition
there confirms the real-space picture: $68$--$97\%$ of $\alpha_{xx}$ is
collective inter-atomic charge flow rather than local atomic dipoles.

A pooled read-out cannot reach this regime. Summing local contributions frozen
beyond $R_c$ locks the size-scaling to strictly extensive ($\tau=1$) or
intensive ($\tau=0$), so a super-linear collective response is structurally out
of reach; even charge-weighted (charge\,$\times$\,position) dipole read-outs
remain capped at linear, and only genuinely non-local schemes (global attention,
charge equilibration) escape the bound, but must then \emph{learn} the size
dependence from data they lack at large $N$ (SI). This is exactly the ablation
of Fig.~\ref{fig:scaling}a: no amount of training data lifts it, because the
missing ingredient is the global eigenproblem, not a richer local
representation.

\FloatBarrier
\section{Discussion}
\label{sec:discussion}

The results show that a single equivariant model can place a broad suite of
molecular properties at the favourable corner of the accuracy--cost
plane, coupled-cluster accuracy, at the cost of a single DFT single
point, across the main group. These results rest on the two principles set out in the Introduction. First,
the \emph{level of supervision} sets the accuracy ceiling: HARP is trained
on the observables, matched to a correlated (CCSD(T)) reference, with the cheap
B3LYP/def2-SVP single point serving only as a baseline the network corrects (so
it need only represent the smooth, relatively small difference between hybrid
DFT and coupled cluster). Because it never fits a reference Hamiltonian matrix,
its accuracy is bounded by the reference level, not by DFT, which is what
separates it from Hamiltonian-regression surrogates. Second, the \emph{architecture}
sets the size-scaling: the network outputs its correction as a local change to
the \emph{Hamiltonian}, and every property is read off by the exact
quantum-mechanical operator acting on the eigenstates of that Hamiltonian. The
learnable part (the Hamiltonian correction $\Delta\mathbf{H}_\theta$) is strictly local and depends only on
chemical environments that recur across sizes and chemistries (which is why
per-element accuracy does not degrade, and why the model transfers from
$\lesssim$25-atom training molecules to 58-atom chains), but the \emph{property}
is computed globally, by diagonalization and linear response, so its
dependence on system size is fixed by quantum mechanics rather than by a
pooling choice. An architecture that instead pools per-atom contributions can only
produce strictly extensive (sum) or strictly intensive (mean) scaling; it
cannot represent a collective response that grows super-linearly with length,
and, because that growth lies outside the small-molecule training
distribution, no amount of data can supply it. Correct size-scaling is, in
this sense, an architectural property, not a learnable one.

The oligothiophene case makes this concrete. The network does not merely echo
its baseline but actively removes the DFT delocalization error, tracking
finite-field CCSD polarizability to $\sim$2\% out to the 44-atom CCSD
limit, as accurately at $\mathrm{T}_6$ as in-distribution, and continuing the
corrected super-linear trend beyond it, where even one CCSD field point costs
$5.7$~h ($\sim$500$\times$ the model's entire inference; Supplementary
Fig.~S5). It can do so because
the longitudinal response of these chains is dominated by collective,
delocalized charge transfer (Supplementary Fig.~S3; 68--97\% of $\alpha_{xx}$),
which the global linear response of the predicted Hamiltonian captures even
though the learned correction is local, exactly the contribution a pooling or
induced-dipole surrogate would miss. A read-out ablation makes this concrete:
a model with the same backbone architecture and dataset but a direct
property read-out drifts back toward the over-polarized baseline as the chain grows, while the
Hamiltonian read-out does not (Fig.~\ref{fig:scaling}). The model is strongest
where its targets
are clean, single-reference coupled-cluster properties: total energies,
polarizabilities, atomic charges, and bond orders. The clear weak link is the
optical gap, whose label is an EOM-CCSD/cc-pVDZ excitation; the model
reproduces that target faithfully even out of distribution, but the target
itself is not a physical gold standard for conjugated systems, and the model
inherits its $\sim$1~eV offset from experiment. This is a property of the
labels, not the surrogate, and it points directly to how the dataset should
evolve (see below).

Practically, the method enables coupled-cluster-quality prediction of the
six ground-state and response properties at high throughput for the nine
main-group elements covered, including the P/S/Cl space relevant to
ligands, agrochemicals, flame retardants, and sulfur- and phosphorus-based
materials, at a cost dominated by a single cheap DFT call. The optical
gap is the exception: it tracks its EOM-CCSD/cc-pVDZ label
faithfully but inherits that label's offset from experiment, so the gap
should be treated as a relative, label-consistent quantity rather than a
substitute for a higher-level spectroscopic prediction. A GPU-accelerated
DFT baseline would reduce the residual cost further, but the limiting
accuracy is set by the labels, not the surrogate.

The model's accuracy is bounded by the level of theory of its training
labels, and our analysis identifies the optical gap as the binding
constraint: its EOM-CCSD/cc-pVDZ label is not a gold standard for
$\pi$-conjugated systems, because the small basis lacks diffuse functions,
the method omits triples, and the relevant excited states acquire
double-excitation character that grows with chain length
~\cite{loos2019doubles,watson2012excited}. The model
reproduces this label faithfully, so improving the gap requires improving the
labels, not the network.

This motivates a concrete dataset roadmap. (i) Add diffuse functions
(aug-cc-pVDZ or def2-TZVPD) for the response and excited-state targets, the
single largest and cheapest accuracy gain; on thiophene this alone recovers
$0.41$~eV of the gap offset. (ii) Make the fundamental gap (IP$-$EA),
obtainable from the bt-PNO-IP/EA-EOM-CCSD data we already compute, the headline gap:
it is a single-electron process, free of the double-excitation pathology, and
robust with system size. (iii) Build out-of-distribution holdouts and a
graded series of conjugated molecules \emph{by design}, so transfer is
measured rather than discovered post hoc. (iv) Calibrate against a
higher-level reference (CC3/aug-cc-pVTZ) on a few hundred molecules to pin
the residual label error, rather than recomputing the entire set at that
level. Full coupled-cluster triples for excitations
(CC3/CCSDT, $\mathcal{O}(N^{7\text{--}8)})$) are intractable for systems of
this size, which is why we bound the gap ceiling with a basis-set study and
literature benchmarks (SI) rather than with explicit triples.

Two further limitations are worth naming explicitly. First, the
training/validation/test split is by shuffled dataset-index range (Methods) rather
than by scaffold or size; while every molecule appears in exactly one set,
this does not guarantee absence of close chemical near-neighbours between
train and test, and scaffold-split benchmarks would more conservatively
measure transfer to genuinely unseen chemistry. Second, our comparison against
other machine-learning models is necessarily heterogeneous: we map released
checkpoints onto the benchmark's chemistry and property suite
(Table~\ref{tab:mlcompare}) and report accuracy for the machine-learned
functional Skala alongside conventional functionals
(Table~\ref{tab:mae}), but the released models differ in training set,
reference level, and target chemistry, so these are not like-for-like error
comparisons. We cleanly isolate the role of the read-out itself through the
read-out ablation (direct vs.\ Hamiltonian read-out;
Fig.~\ref{fig:scaling}); a controlled retraining of competing architectures
on our common dataset and CCSD(T) reference would extend this to a full
head-to-head accuracy benchmark and is left to future work.

\FloatBarrier
\section{Methods}
\label{sec:methods}

\subsection{Dataset and reference levels}
The dataset comprises $42{,}912$ closed-shell, neutral main-group molecules
(2--24 atoms, $\le 9$ heavy) built from H, C, N, O, F, Si, P, S, and Cl,
retained after quality control and outlier removal from $44{,}412$ molecular
structures collected from PubChem (one 3D structure per compound; SI). The
molecules are partitioned by dataset-index range into training ($40{,}985$;
indices $2001$--$44{,}412$), validation ($968$; $1001$--$2000$) and test
($959$; $1$--$1000$).

Each molecule appears in exactly one set. Because the structures were pseudorandomly
shuffled and relabeled $1$--$44{,}412$ before any reference calculations (SI),
this index-range split is random with respect to chemistry and molecule size; it is, however, \emph{not} scaffold-split or
size-stratified, and we therefore do not guarantee absence of close
chemical near-neighbours between train and test, a limitation that the
out-of-distribution oligothiophene benchmark (Fig.~\ref{fig:scaling})
is partly designed to address. Each property is labeled at a
property-appropriate level. The electronic energy---and likewise the dipole,
quadrupole, Mulliken atomic charges, and Mayer bond orders, evaluated on the
CCSD one-particle density of each run (linearized for canonical CCSD(T),
unrelaxed for DLPNO-CCSD(T); SI)---is a \emph{composite} estimate of canonical
CCSD(T)/cc-pVTZ, assembled from three coupled-cluster runs per molecule as
\begin{equation}
X_{\mathrm{label}} \;=\; X^{\text{cc-pVDZ}}_{\mathrm{CCSD(T)}}
\;+\; X^{\text{cc-pVTZ}}_{\mathrm{DLPNO}}
\;-\; X^{\text{cc-pVDZ}}_{\mathrm{DLPNO}},
\label{eq:composite}
\end{equation}
i.e., a canonical CCSD(T)/cc-pVDZ value plus a
DLPNO-CCSD(T)~\cite{riplinger2013dlpno} basis-set correction from cc-pVDZ to
cc-pVTZ; we write ``composite CCSD(T)/cc-pVTZ'' for this level throughout.
The isotropic and full
polarizability is labeled from finite-field CCSD/cc-pVDZ (central finite
differences of the CCSD dipole moment in static fields of $\pm10^{-4}$~a.u.\
along each axis, seven calculations per molecule);
the lowest singlet vertical excitation (the gap $E_g$) from
EOM-CCSD/cc-pVDZ~\cite{stanton1993eom} (lowest root); and ionization
potentials and electron affinities are from bt-PNO-IP/EA-EOM-CCSD/cc-pVDZ. Two
basis-set-dependence caveats apply: (i) the polarizability reference is
cc-pVDZ \emph{without} diffuse functions and is therefore itself biased
relative to a complete-basis CCSD limit by a few percent (SI), so
``$\sim$2\% of CCSD'' should be read as ``$\sim$2\% of this systematically
biased reference''; (ii) Mulliken charges and Mayer bond orders are
basis-dependent partitioning schemes: the B3LYP/def2-SVP baseline and the
composite CCSD(T)/cc-pVTZ target are strictly different quantities, and the model
learns the correction between them rather than predicting an absolute
basis-independent ``physical'' partial charge. The $\Delta$-learning
baseline (and the baseline Fock matrix $\mathbf{H}_{\mathrm{DFT}}$ that the
network corrects) is a single B3LYP/def2-SVP calculation. All
reference calculations were performed with ORCA~6.0~\cite{neese2022orca};
protocols and input templates are in the SI.

\subsection{Model architecture}
Rather than reading each target property out of pooled atomic features, the
model predicts an \emph{effective one-electron Hamiltonian} and derives every
property from it by the same quantum-mechanical operator that defines it in
the reference calculation. This extends the molecular-Hamiltonian-learning
formulation of Tang \emph{et al.}~\cite{tang2024multitask} (demonstrated
there for hydrocarbons) to the nine-element main group, and is the structural
source of the model's size transferability (Fig.~\ref{fig:scaling},
Discussion).

The backbone is an SO(3)-equivariant, spherical-harmonic graph-attention
network (EquiformerV2~\cite{liao2024equiformerv2}, in the eSCN
lineage~\cite{passaro2023escn}): four interaction layers, a radial cutoff of
$6.0$~\AA{} with up to $20$ neighbours per atom, $64$ spherical channels,
$8$ attention heads, spherical-harmonic degree $l_{\max}=m_{\max}=4$---the
degree required to represent the $d\times d$ on-site Fock blocks that the
third-row elements (Si, P, S, Cl) introduce---and RMS
spherical-harmonic normalization with gate activations ($\sim$5.6~M
parameters). Its equivariant node features are mapped by
e3nn~\cite{geiger2022e3nn} heads, per element for on-site blocks and pairwise
(combined with edge spherical harmonics) for off-site blocks, onto a
correction $\Delta\mathbf{H}_\theta$ to the baseline
Fock matrix in the atomic-orbital basis. The effective Hamiltonian is
$\mathbf{H}=\mathbf{H}_{\mathrm{DFT}}+\Delta\mathbf{H}_\theta$, where
$\mathbf{H}_{\mathrm{DFT}}$ is the B3LYP/def2-SVP Kohn--Sham matrix of the
cheap baseline single point; solving the eigenvalue equations of $\mathbf{H}$
yields molecular-orbital energies $\{\varepsilon_n\}$ and coefficients, hence
the one-particle density matrix.

Every property is then evaluated from $\mathbf{H}$ and its eigenstates by the
defining operator, not by a separate learned head: the total electronic energy
as the sum over occupied levels; the gap as the frontier (HOMO--LUMO)
eigenvalue difference; the dipole and quadrupole as the multipole operators
traced against the density; the polarizability by linear response (detailed below); and Mulliken atomic
charges and Mayer bond orders from the density matrix.

The polarizability is built in two steps from the eigenpairs
$\{\varepsilon_n,\psi_n\}$ of $\mathbf{H}$. A bare independent-particle
sum-over-states response,
\begin{equation}
\alpha^{0}_{xy} = 4\!\!\sum_{i\in\text{occ}}\;\sum_{a\in\text{virt}}
\frac{\langle\psi_i|\,\hat{x}\,|\psi_a\rangle\,
      \langle\psi_i|\,\hat{y}\,|\psi_a\rangle}
     {\varepsilon_a-\varepsilon_i},
\label{eq:sos}
\end{equation}
is screened (depolarization) by the screening matrix
$\mathbf{T}=\mathbf{T}_{\mathrm{DFT}}+\Delta\mathbf{T}_\theta$ to give the reported
tensor,
\begin{equation}
\boldsymbol{\alpha} =
\bigl(\mathbf{I}+\boldsymbol{\alpha}^{0}\mathbf{T}\bigr)^{-1}\boldsymbol{\alpha}^{0},
\label{eq:screen}
\end{equation}
where the baseline screening matrix $\mathbf{T}_{\mathrm{DFT}}$ is obtained from a
linear-response (TDDFT) calculation on the baseline
(so an uncorrected model reproduces the baseline polarizability; this TDDFT
step is needed only when the polarizability is requested)
and $\Delta\mathbf{T}_\theta$ is the learned correction. Because the read-out is the exact quantum-mechanical functional
of a Hamiltonian assembled from a \emph{local}, finite-cutoff correction,
intensive and extensive properties inherit their correct system-size
dependence \emph{by construction}, rather than from a pooling choice (sum,
which forces strict extensivity, or mean, which forces strict intensivity).
This is what allows the model to extrapolate the size-dependence of collective
properties beyond the training set (Fig.~\ref{fig:scaling}). With the B3LYP
baseline the Kohn--Sham gap already tracks the correlated optical gap with
near-unit slope, so, unlike a GGA baseline, no multiplicative gap rescaling
is needed.

\subsection{Training}
The model was trained on the full $40{,}985$-molecule training partition by
distributed data parallelism across 16 NVIDIA A100 GPUs (4 nodes), minimizing
a weighted sum of per-property mean-squared-error losses (the per-property
weights balance the disparate property scales; values in the released
config), using the Adam optimizer~\cite{kingma2015adam} with a step-decay
schedule taking the learning rate from $3\times10^{-3}$ to $10^{-4}$ over the
$2{,}000$-epoch schedule. Regularization comprised attention dropout ($0.1$)
and stochastic depth (drop-path $0.05$). Each epoch is one pass over the
training partition in $100$ minibatches ($\sim$410 molecules per global
batch); training ran for $1{,}375$ epochs ($1.4\times10^{5}$ gradient steps,
$\sim$50~h of wall-clock time across resumable six-hour allocations), by which
point the validation loss had plateaued; the epoch-1265 checkpoint, which
minimizes the validation loss, is used throughout.
Full hyperparameters are in the released configuration file.

\paragraph{Training stability and level tracking.}
A Hamiltonian read-out introduces a failure mode absent from property-head
models: as the learned correction $\Delta\mathbf{H}_\theta$ shifts the spectrum
during training, two molecular orbitals can cross, and a strict energy
ordering then reassigns which orbital is the HOMO discontinuously, producing a
jump in the density, and in every response property derived from it, that
destabilizes the gradient. We addressed this with a diabatic
\emph{level-tracking} scheme: at each step the occupied subspace is identified
not by the $n_e$ lowest eigenvalues but by maximal overlap of each eigenstate
with the baseline occupied density, so the occupied/virtual assignment varies
continuously through a crossing. This is distinct from, and complementary to,
the perturbation-theory treatment of near-degenerate eigenvalue
\emph{gradients}~\cite{tang2024multitask}: it removes the discontinuity in the
\emph{forward} assignment, not the singularity in the backward pass. Level
tracking was essential for stable training on the unfiltered crawl; once the
dataset was quality-filtered (Methods, SI), surviving crossings were rare
enough that plain energy ordering sufficed, and the released model dispenses
with it. We document the scheme because it is a cheap, general remedy for
training Hamiltonian-prediction models on noisier data.

\subsection{Benchmark systems and timing}
Oligothiophene geometries were generated
programmatically (planar, idealized; SI). Reference properties on these
geometries used ORCA~6.0~\cite{neese2022orca}: analytic BP86 polarizabilities
via the ELPROP module, finite-field CCSD/cc-pVDZ polarizabilities (7-point
central differences), EOM-CCSD/cc-pVDZ excitations (lowest three roots), and
BP86 and B3LYP single points (the latter is the model's baseline). The
distributed-polarizability decomposition (Supplementary Fig.~S3) partitions the
analytic finite-field DFT response density via the ELPROP module. Neural-inference
timings (network forward pass plus Hamiltonian diagonalization, screened
response, and property evaluation) were measured on a single NVIDIA A100
(40~GB) with batch size 1, five warmup passes, and the minimum of three timed
passes per molecule;
DFT-baseline wall-times are the ORCA \texttt{TOTAL RUN TIME} on 7 CPU cores at
dataset-generation time.

\section*{Supplementary Information}
The Supplementary Information is appended to this document, covering: the dataset (composition, reference levels, quality control,
single-reference diagnostics); per-property, per-element, error-distribution
and polarizability-tensor accuracy tables; the reference-baseline ablation; the
size-scaling mechanism with frontier-orbital
measurements and a charge-transfer decomposition; the basis-set ceiling of the
optical-gap reference (cc-pVDZ vs.\ aug-cc-pVDZ); the full inference-timing
analysis; the architecture comparison with MEHnet; a scope map of released ML
models; and computational protocols.

\section*{Data availability}
The $959$-molecule held-out test set---the three-dimensional structures
together with every coupled-cluster label reported here (composite
CCSD(T)/cc-pVTZ, finite-field CCSD/cc-pVDZ and EOM-CCSD/cc-pVDZ)---underlies
all quantitative results in this work and is openly available in a Zenodo
archive whose DOI will be minted upon publication. The same record provides
the train/validation/test split definitions, the
three-dimensional structures of all $44{,}412$ source molecules, and their
PubChem identifiers ($97.7\%$ resolved against current PubChem; each
remaining entry is classified, with coordinate-level evidence, as delisted
since the crawl). The full corpus is therefore reconstructable from the
record and the released ORCA input templates alone (see Code availability),
with no further dependence on PubChem. Of the $42{,}912$-molecule
train/validation/test partition, the coupled-cluster labels of the training and
validation sets are not redistributed with that record; they can be regenerated from the released structures and
input templates, and are available from the corresponding author on reasonable
request.

\section*{Code availability}
The model implementation, the training and inference pipeline, the trained
HARP weights (release v1.0), the ORCA input templates used to generate every
reference label and baseline, and the quality-control and filtering code are
openly available at \url{https://github.com/He-Wenhao/HARP} under the MIT
license; the release will be archived on Zenodo upon publication.

\section*{Acknowledgements}
This work was supported by the Honda Research Institute. N.S. and Y.W. acknowledge support from U.S. Department of Energy, Office of Science, Basic Energy Sciences, under Early Career Award No.~DE-SC0024524. This research used
resources of the National Energy Research Scientific Computing Center (NERSC), a
U.S.\ Department of Energy Office of Science User Facility at Lawrence Berkeley
National Laboratory, operated under Contract No.~DE-AC02-05CH11231 (the
Perlmutter system), which provided the primary computational resources for this
work. Additional computing resources were provided by the Frontera system at the
Texas Advanced Computing Center (TACC) at The University of Texas at Austin.

\section*{Author contributions}
W.H.\ designed and implemented the model, performed the experiments and
analysis, and wrote the manuscript. H.T.\ contributed to the
methodology. X.C.\ implemented the distributed-training parallelization and
optimized GPU utilization. N.S.\ generated the training dataset. T.S.H. optimized observable back-propagation pipelines. Z.L.\ and B.L.\ studied the alignment of
the model with experiment. Y.Y.\ contributed the neural-network training
methodology. F.L., Y.W.\ and J.L.\ provided
computational resources, supervised the project, acquired funding, and edited
the manuscript. A.R.H.\ co-initiated the research theme, co-formulated the research goals, and acquired funding. All authors discussed the results and commented on the
manuscript.

\section*{Competing interests}
The authors declare no competing interests.

\bibliographystyle{unsrt}
\bibliography{refs}

\clearpage\onecolumn


\section*{Supplementary material}

# S1 The dataset

The model is trained on a purpose-built, multi-property, coupled-cluster reference dataset that we describe here at the level of detail of a stand-alone data resource. Each molecule carries a full suite of property labels computed at property-appropriate correlated levels, on top of cheap DFT baselines, and the set passes a documented multi-stage quality-control protocol. Because the training-partition labels are not redistributed (Section S1.7), this section is intended to be self-contained.

## S1.1 Scope and composition

The dataset comprises 42,912 neutral, closed-shell singlet molecules spanning nine main-group elements (H, C, N, O, F, Si, P, S, Cl), retained after the quality filter of Section S1.4 from 44,412 crawled structures. The molecules are small (2–24 atoms, $\leq$ 9 heavy atoms; median 14 atoms, 7 heavy), which fixes the size bound for the out-of-distribution size-extrapolation tests of the main text—bithiophene ($T_2$), with 10 heavy atoms, already exceeds it. The distinguishing feature is element coverage beyond the organic C/H/N/O core: sulfur appears in 58.1% of molecules, nitrogen in 56.8%, oxygen in 47.6%, chlorine in 21.9%, phosphorus in 15.2%, silicon in 9.5%, and fluorine in 6.5% (Section S1.6). The substantial S, P, and Cl representation—not a handful of token molecules—is what supports the main-group accuracy claims.

**Molecular structures.** Geometries were crawled from PubChem (one three-dimensional structure per compound), populating phosphorus-, sulfur-, and chlorine-containing chemistry alongside the organic elements. These are the standard PubChem computed 3D conformers (near-equilibrium structures); no further geometry optimization or conformer sampling was applied before the reference calculations. Before any reference calculation the structures were pseudorandomly shuffled and relabeled with sequential indices 1–44,412 (replacing the native PubChem identifiers); these dataset indices define the train/validation/test partition, so although the split is specified by index range it is random with respect to chemistry and molecule size (Section S1.4).

## S1.2 Multi-property reference levels

Every geometry was characterized with a fixed stack of single-point calculations in ORCA 6.0 [1], and each property is taken from the level appropriate to it (Table S1). Two DFT single points (BP86/def2-SVP and B3LYP/def2-SVP) are computed for every molecule and serve as $\Delta$-learning baselines: the released model corrects the *B3LYP* Fock matrix, while the BP86 baseline supports the reference ablation of Section S3. DSD-PBEP86-D3(BJ)/def2-TZVP (frozen core, RIJK and RI-MP2) is computed as a double-hybrid comparator. Input templates for every method are released with the code.

The ground-state observables are composites of three coupled-cluster runs: canonical CCSD(T)/cc-pVDZ plus a DLPNO-CCSD(T) basis-set correction from cc-pVDZ to cc-pVTZ (main-text Methods). The DLPNO-CCSD(T) steps use RIJCOSX with the cc-pV$X$Z/C correlation-fitting and

Table S1: **Reference levels per property.** All calculations use ORCA 6.0.0. The two DFT baselines (BP86, B3LYP) are def2-SVP single points; the double-hybrid comparator uses def2-TZVP. RIJCOSX (def2/J, cc-pV$X$Z/C) is used only in the DLPNO-CCSD(T) steps; the canonical CCSD(T) and CCSD runs use no density fitting.

| quantity | level of theory |
|---|---|
| $\Delta$-learning baseline (released model) | B3LYP/def2-SVP |
| $\Delta$-learning baseline (ablation, Sec. S3) | BP86/def2-SVP |
| DFT comparator | DSD-PBEP86-D3(BJ)/def2-TZVP |
| energy, multipoles, Mulliken charges, Mayer bond orders | composite CCSD(T)/cc-pVTZ (see below) |
| polarizability $\alpha_{ij}$ | finite-field CCSD/cc-pVDZ (dipole derivatives) |
| optical gap (lowest singlet excitation) | EOM-CCSD/cc-pVDZ |
| ionization potential / electron affinity | bt-PNO-IP/EA-EOM-CCSD/cc-pVDZ |

def2/J Coulomb-fitting auxiliary bases and TightSCF. The energies are CCSD(T); the electrostatic moments (ELPROP), Mulliken charges, and Mayer bond orders are evaluated on the CCSD one-particle density of each run—the linearized CCSD density for canonical CCSD(T)/cc-pVDZ and the unrelaxed CCSD density (`ENFORCE_CCSD_DENSITY`) for the two DLPNO steps—without orbital relaxation. Polarizabilities are finite-field dipole derivatives: the full tensor $\alpha_{ij} = [\mu_i(+F_j) - \mu_i(-F_j)]/(2F)$ is built from CCSD/cc-pVDZ dipole moments (linearized density) in static fields of $\pm F$ along each axis, $F = 1 \times 10^{-4}$ a.u., i.e. seven CCSD calculations per molecule (ORCA compound script). The optical-gap label is the lowest of three EOM-CCSD/cc-pVDZ singlet roots (canonical, TightSCF). The bt-PNO-IP/EA-EOM-CCSD/cc-pVDZ ionization potentials and electron affinities are stored for every molecule; their difference (the fundamental gap IP$-$EA, median 6.4 eV across the set) is a single-electron, size-robust alternative to the optical gap that the dataset already supports (main-text Discussion roadmap).

## S1.3 Generation pipeline and throughput

The pipeline is a single ORCA stack run per geometry, parsed into one per-molecule record holding the baseline Fock matrices and all correlated labels. The dominant cost is the correlated stack (the three-run composite CCSD(T) stack for the ground-state observables, seven finite-field CCSD/cc-pVDZ runs for the response, one EOM-CCSD/cc-pVDZ solve for the excitations, and two bt-PNO-EOM-CCSD/cc-pVDZ solves for the IP and EA); the DFT baselines are negligible by comparison. Keeping the system size small ($\leq 9$ heavy atoms) is what makes the correlated labels affordable at the $\sim$44k scale. The full protocol (methods, bases, fitting sets, convergence settings, and finite-field parameters) is released as the ORCA input templates of the production runs, so the dataset can be regenerated or extended.

## S1.4 Quality control and filtering

The reference calculations are extensive and some fail to converge or yield unphysical values; we therefore filter them in two steps (Table S2).

*Calculation-level quality control.* When parsing raw ORCA outputs into per-molecule records, a molecule is discarded if any required calculation failed to converge or could not be parsed (e.g. the CCSD(T), finite-field CCSD, or EOM-CCSD step did not complete), or if the lowest EOM-CCSD excitation energy is missing or negative—an unphysical "gap" that signals a failed or root-flipped excited-state solve.

*Model-in-the-loop outlier removal.* Residual label errors (corrupt reference densities that nonetheless parsed) were caught self-consistently: we trained a BP86-baseline model, evaluated it on every molecule with a usable BP86 baseline (41,948 of those passing quality control), and computed the per-molecule squared error of each property. For each property we formed a robust $z$-score on $\log_{10}$(error) using the median and median absolute deviation ($\sigma = 1.4826\,\mathrm{MAD}$), and flagged any molecule with $z > 10$ on any property. This removed 9 molecules (dominated by anomalous polarizability and quadrupole values, e.g. a predicted-vs-reference $\alpha$ discrepancy of $> 10^4$ a.u.$^2$, diagnostic of a corrupt reference rather than model error), whose records were deleted from the dataset. The two steps leave 42,912 molecules, the dataset used throughout this work.

The remaining 973 quality-controlled molecules lack a usable BP86 baseline and so were not part of this outlier screen. Their coupled-cluster labels pass the same quality control and they have valid B3LYP baselines, so they are retained for the released (B3LYP-baseline) model. A model built on the BP86 baseline skips them at load time; this affects only the BP86-baseline reference ablation (Section S3).

Table S2: **Data filtering.** Molecules retained after quality control and model-in-the-loop outlier removal.

| stage | criterion | retained |
|---|---|---|
| crawled | 3D structures crawled from PubChem | 44,412 |
| QC + outliers | CCSD(T)/EOM converged, parseable, gap > 0; | |
| | robust log-error $z \leq 10$ on every property | 42,912 |

The retained molecules are partitioned by index range into training (40,985; indices 2001–44,412), validation (968; 1001–2000), and test (959; 1–1000) sets, disjoint by construction. Because the structures were shuffled before any calculation, this index-range split is random with respect to chemistry and size.

### S1.5 Domain of validity: single-reference diagnostics

CCSD(T) is the gold standard only in the single-reference regime, so the dataset is reliable as a reference only to the extent that its molecules are single-reference. We therefore record the $T_1$ diagnostic for every molecule (the coupled-cluster $T_1$ amplitude norm, not to be confused with the oligothiophene labels $\mathrm{T}_n$ used in later sections). Across all 42,912 molecules the $T_1$ distribution is tight and well below the conventional single-reference threshold of 0.02 (Table S3): the median is 0.012, the 99th percentile 0.019, and no molecule exceeds 0.020. The dataset thus lies entirely within the domain where the coupled-cluster labels are trustworthy, consistent with its composition of small, neutral, closed-shell main-group molecules at near-equilibrium geometries. Static correlation (diradicals, bond-breaking, near-degeneracy) is by construction out of scope, and CCSD(T) reference quality should not be assumed there.

Table S3: $T_1$ single-reference diagnostic over the 42,912-molecule set. All molecules satisfy $T_1 \leq 0.02$, the standard threshold below which single-reference coupled cluster is reliable.

| | median | mean | $p_{90}$ | $p_{99}$ | max |
|---|---|---|---|---|---|
| $T_1$ | 0.0122 | 0.0126 | 0.0163 | 0.0190 | 0.0200 |

## S1.6 Dataset statistics

Figure S1 and Table S4 characterize the full 42,912-molecule dataset (596,239 atoms in total). The molecules contain between 2 and 24 atoms (median 14, mean 13.9) and between 1 and 9 heavy (non-hydrogen) atoms (median 7).

Table S4: Per-element statistics over the 42,912-molecule dataset: number of atoms, number of molecules containing the element, and that fraction of the dataset.

| element | H | C | N | O | S | Cl | P | Si | F |
|---|---|---|---|---|---|---|---|---|---|
| atoms | 310,457 | 167,376 | 36,386 | 29,967 | 26,645 | 9,992 | 7,071 | 4,637 | 3,708 |
| molecules | 42,600 | 42,101 | 24,381 | 20,425 | 24,933 | 9,389 | 6,517 | 4,074 | 2,790 |
| % of dataset | 99.3 | 98.1 | 56.8 | 47.6 | 58.1 | 21.9 | 15.2 | 9.5 | 6.5 |

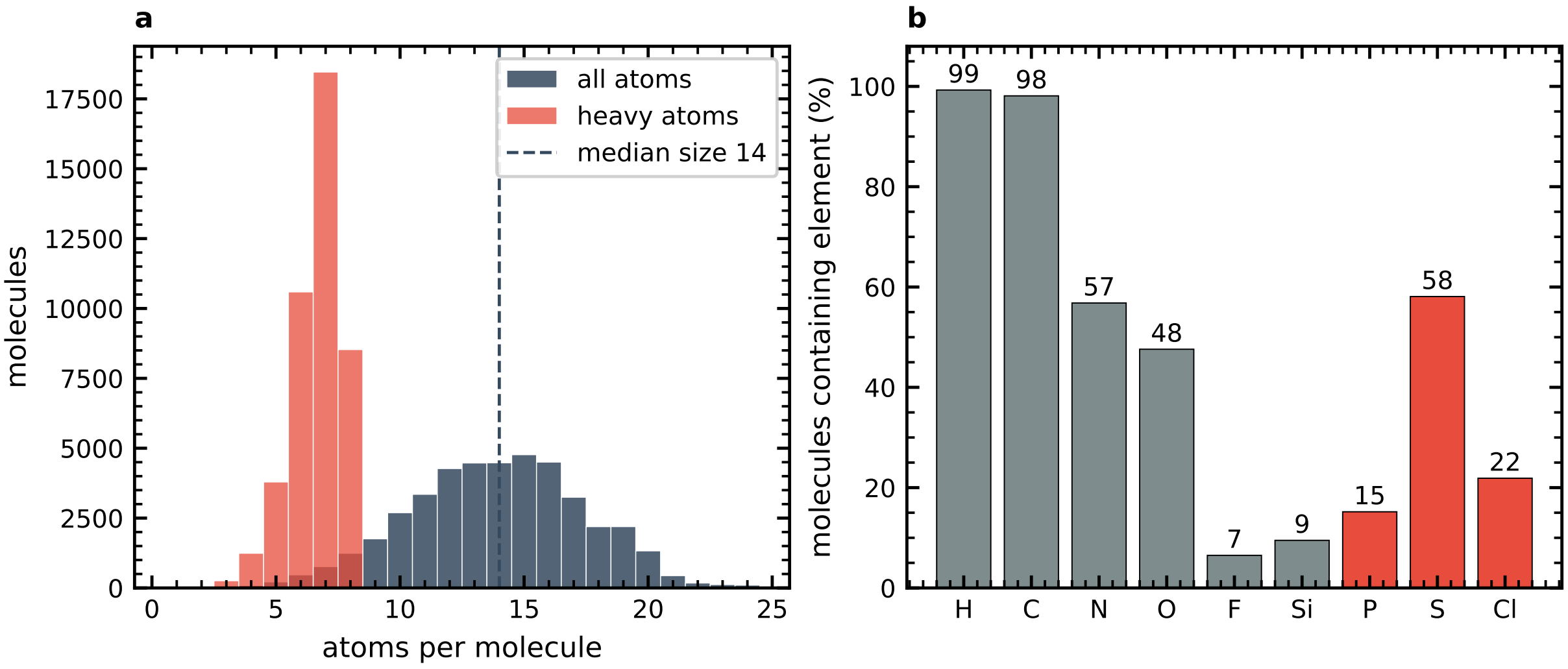


Figure S1: Dataset statistics. **(a)** Distribution of molecule size (all atoms and heavy atoms); molecules span 2–24 atoms ($\leq$ 9 heavy). **(b)** Fraction of molecules containing each element; the P/S/Cl elements (highlighted) are well represented, not token cases.

## S1.7 Data and code availability

The 959-molecule test set (structures and all coupled-cluster labels), the train/validation/test split definitions, the three-dimensional structures of all 44,412 source molecules, and their PubChem identifiers (97.7% resolved against current PubChem; the remainder delisted since the crawl) are openly available; the model implementation, trained weights, ORCA input templates, and the quality-control/filter code are in a public repository. Both records are identified in the main-text Data availability and Code availability statements. The coupled-cluster labels for the training and validation partitions are not redistributed; they can be regenerated from the released structures and input templates, and are available from the corresponding author on reasonable request.

## S2 Extended accuracy benchmark

This section reports the per-property, per-element, and full-distribution accuracy that the main text condenses into a single MAE per property (main-text Table 1), and the full polarizability tensor behind the isotropic value. All numbers are for the released model (HARP, B3LYP baseline, epoch-1265 checkpoint; Section S1.2) on the 959-molecule test set, referenced to the composite CCSD(T)/cc-pVTZ ground truth (main-text Methods), and are the same evaluation as main-text Table 1.

### S2.1 Per-property error

Table S5 reports mean absolute and root-mean-square errors for the model and the three DFT methods, in conventional per-property units. The model improves on the best DFT method for every property: the energy MAE is 0.23 $\text{kcal mol}^{-1}$ versus 13.2 $\text{kcal mol}^{-1}$ for B3LYP, and the response properties improve severalfold to an order of magnitude even relative to the double hybrid. Main-text Table 1 reports the dipole and quadrupole as full-vector and Frobenius-norm errors; here they are per Cartesian component.

Table S5: Per-property error (MAE / RMSE) on the 959-molecule test set, referenced to composite CCSD(T)/cc-pVTZ. As in main-text Table 1, each functional is evaluated at the basis of the property's coupled-cluster reference (cc-pVTZ for $E$, dipole, quadrupole, Mulliken charge and Mayer bond order; cc-pVDZ for $\alpha_{\text{iso}}$ and $E_g$). Best MAE in bold.

| property | unit | this work | BP86 | B3LYP | DSD-PBEP86 |
|---|---|---|---|---|---|
| $E$ | $\text{kcal mol}^{-1}$ | **0.231** / 0.380 | 63.0 / 64.4 | 13.2 / 15.0 | 15.1 / 16.0 |
| $E_g$ | eV | **0.067** / 0.113 | 1.317 / 1.409 | 0.609 / 0.722 | 4.838 / 4.910 |
| dipole | a.u. (per comp.) | **0.0041** / 0.0074 | 0.0291 / 0.0465 | 0.0291 / 0.0463 | 0.0200 / 0.0306 |
| quadrupole | a.u. (per comp.) | **0.0123** / 0.0213 | 0.112 / 0.164 | 0.0899 / 0.169 | 0.0625 / 0.111 |
| $\alpha_{\text{iso}}$ | a.u. | **0.121** / 0.212 | 4.712 / 5.087 | 2.792 / 3.250 | 1.182 / 1.409 |
| Mulliken charge | e (per atom) | **0.0053** / 0.0082 | 0.0205 / 0.0316 | 0.0256 / 0.0375 | 0.0220 / 0.0287 |
| Mayer bond order | per pair (bonded) | **0.0066** / 0.0106 | 0.0556 / 0.0623 | 0.0588 / 0.0656 | 0.0251 / 0.0286 |

### S2.2 Error distributions

Mean-based summaries can hide heavy-tailed behaviour. Table S6 reports the model's absolute-error distribution: median, 90th and 99th percentiles, and worst case on the 959-molecule test set. The distributions are moderately skewed—$p_{99}$ is typically 6–14× the median. For the energy, 99% of test molecules lie within 1.3 $\text{kcal mol}^{-1}$ (median 0.15 $\text{kcal mol}^{-1}$); the single worst case, 5.0 $\text{kcal mol}^{-1}$, is an isolated outlier rather than the edge of a heavy tail. The gap has the broadest relative distribution ($p_{99} \approx 0.50$ eV, max 0.77 eV), consistent with the label-level discussion in the main text.

### S2.3 Full polarizability tensor

The main text reports only the isotropic polarizability $\alpha_{\text{iso}} = \frac{1}{3}\text{tr}\,\boldsymbol{\alpha}$. Table S7 resolves the full tensor into its isotropic part, the diagonal components ($\alpha_{xx}, \alpha_{yy}, \alpha_{zz}$), the off-diagonal components

Table S6: Distribution of the model's absolute error on the 959-molecule test set. Energy in $\mathrm{kcal\,mol^{-1}}$, gap in eV, dipole/quadrupole/$\alpha_{\mathrm{iso}}$ in atomic units, Mulliken charges in $e$ (per atom), Mayer bond orders dimensionless (per bonded pair).

| property | median | $p_{90}$ | $p_{99}$ | max |
|---|---|---|---|---|
| $E$ ($\mathrm{kcal\,mol^{-1}}$) | 0.152 | 0.518 | 1.309 | 5.030 |
| $E_g$ (eV) | 0.036 | 0.167 | 0.502 | 0.773 |
| dipole, $\|\Delta\boldsymbol{\mu}\|$ (au) | 0.0061 | 0.0169 | 0.0485 | 0.0945 |
| quadrupole, $\|\Delta\mathbf{Q}\|_F$ (au) | 0.0340 | 0.0786 | 0.224 | 0.366 |
| $\alpha_{\mathrm{iso}}$ (au) | 0.073 | 0.260 | 0.882 | 1.948 |
| Mulliken charge (e) | 0.0035 | 0.0120 | 0.0292 | 0.112 |
| Mayer bond order | 0.0044 | 0.0142 | 0.0354 | 0.199 |

$(\alpha_{xy}, \alpha_{xz}, \alpha_{yz})$, and the polarizability anisotropy

$$\Delta\alpha = \sqrt{\tfrac{1}{2}\left[(\alpha_{xx} - \alpha_{yy})^2 + (\alpha_{yy} - \alpha_{zz})^2 + (\alpha_{zz} - \alpha_{xx})^2 + 6(\alpha_{xy}^2 + \alpha_{xz}^2 + \alpha_{yz}^2)\right]},$$

which captures the directional response that the isotropic average discards. The model corrects every channel, not just the trace: it is accurate on the diagonal, off-diagonal, and anisotropy channels, between 5× (off-diagonal) and 13× (isotropic) below the B3LYP baseline it corrects. The anisotropy MAE of 0.24 a.u. (vs. 2.2 a.u. for B3LYP) confirms that the model captures the shape of the response tensor, not merely its size. The baseline columns here are the def2-SVP responses that enter the model; they therefore differ from the cc-pVDZ functional values of Table S5.

Table S7: Polarizability-tensor error (MAE / RMSE, a.u.) on the 959-molecule test set, by tensor channel, for the model and the two def2-SVP DFT baselines it can be built on. "diagonal" and "off-diagonal" aggregate the three respective components per molecule; $\Delta\alpha$ is the anisotropy.

| channel | this work | B3LYP | BP86 |
|---|---|---|---|
| isotropic $\alpha_{\mathrm{iso}}$ | **0.121** / 0.212 | 1.57 / 2.10 | 3.37 / 3.81 |
| diagonal ($\alpha_{xx,yy,zz}$) | **0.162** / 0.312 | 1.80 / 2.75 | 3.45 / 4.56 |
| off-diagonal ($\alpha_{xy,xz,yz}$) | **0.082** / 0.148 | 0.452 / 0.721 | 0.486 / 0.764 |
| anisotropy $\Delta\alpha$ | **0.244** / 0.432 | 2.19 / 3.31 | 3.26 / 4.72 |

## S2.4 Per-element accuracy

Table S8 breaks the model's atomic-charge error down by element. Accuracy is uniform across all nine elements and, crucially, the rare elements are not systematically worse: fluorine (85 atoms in the test set) and silicon (86 atoms) are predicted about as accurately as the abundant C/H/N/O (Si 0.0094 e versus C 0.0087 e), indicating the model has learned transferable per-element chemistry rather than exploiting the statistics of the common organic motifs. This is the per-element drilldown underlying main-text Fig. 2c.

**Equivariance.** The architecture and its vector and rank-2 tensor heads are $E(3)$-equivariant by construction, so rotating or translating the input geometry rotates/translates the predicted dipole, quadrupole, and polarizability accordingly and leaves the scalar predictions invariant; we verified

Table S8: Model atomic-charge MAE (e) per element on the test set, with the number of atoms of each element.

| element | H | C | N | O | F | Si | P | S | Cl |
|---|---|---|---|---|---|---|---|---|---|
| MAE (e) | 0.0031 | 0.0087 | 0.0074 | 0.0049 | 0.0043 | 0.0094 | 0.0105 | 0.0061 | 0.0043 |
| $N$ atoms | 6931 | 3789 | 814 | 666 | 85 | 86 | 164 | 619 | 208 |

numerically that rotation-of-output and output-of-rotation agree to floating-point precision over random rotations of the test molecules.

## S3 Reference ablation: choice of $\Delta$-learning baseline

The model corrects a cheap DFT baseline rather than predicting properties from scratch, so a natural question is how much the result depends on which baseline functional is used. To isolate this choice we trained two otherwise identical models—same EquiformerV2 backbone, hyperparameters, split, and a common ~10k-molecule subset of the training partition—that correct a *BP86/def2-SVP* (GGA) or a *B3LYP/def2-SVP* (hybrid) baseline. The subset keeps the pair affordable to train; both columns of Table S9 are therefore less accurate than the released model, which uses the B3LYP baseline and the full training partition (Table S5). The one deliberate difference concerns the gap: the BP86 variant keeps the learned affine gap rescaling of MEHnet (Section S6), which a GGA Kohn–Sham gap requires, whereas the B3LYP variant reads the gap directly off the frontier eigenvalues. Both are evaluated on the 959-molecule test set against the same CCSD(T) ground truth.

The two baselines give essentially the same accuracy: every property agrees within a factor of ~1.5, and both lie well below the DFT functionals of Table S5, improving on the best functional for each property by a factor of ~2.5–3 (Mulliken charges, multipoles) to ~25 (energy). The framework is thus not tied to a particular baseline functional: it corrects whatever single point it is given onto the coupled-cluster target. The one systematic difference is the optical gap, where the B3LYP baseline is clearly better (0.106 vs. 0.145 eV) *despite* using no learned rescaling. This is the reason the released model uses B3LYP: its Kohn–Sham gap already tracks the correlated optical gap closely enough that the gap can be read directly off the predicted Hamiltonian, removing the pooled rescaling head whose size behaviour is not guaranteed (Section S6). The BP86 baseline reaches comparable energy and response accuracy but pays for it on the gap; B3LYP buys the cleaner, rescaling-free gap read-out at no cost to the other properties.

Table S9: **Reference ablation.** Per-property error (MAE / RMSE) of two otherwise identical models, trained on the same ~10k-molecule training subset, that correct a B3LYP versus a BP86 baseline, on the 959-molecule test set against composite CCSD(T)/cc-pVTZ. Both lie far below the DFT functionals (Table S5); the framework is reference-agnostic except for the gap, where the hybrid baseline is preferred. The released full-data model is reported in Table S5.

| property | unit | B3LYP baseline | BP86 baseline |
|---|---|---|---|
| $E$ | kcal mol$^{-1}$ | 0.515 / 0.768 | 0.540 / 0.769 |
| $E_g$ | eV | 0.106 / 0.162 | 0.145 / 0.208 |
| dipole | a.u. (per comp.) | 0.0074 / 0.0128 | 0.0059 / 0.0100 |
| quadrupole | a.u. (per comp.) | 0.0214 / 0.0353 | 0.0184 / 0.0298 |
| $\alpha_{\text{iso}}$ | a.u. | 0.203 / 0.365 | 0.208 / 0.417 |
| Mulliken charge | e (per atom) | 0.0080 / 0.0122 | 0.0066 / 0.0106 |
| Mayer bond order | per pair (bonded) | 0.0099 / 0.0153 | 0.0086 / 0.0134 |

## S4 Out-of-distribution size extrapolation

This section gives the size-scaling argument condensed in the main text (Results, "Out-of-distribution size scaling and its mechanism"), the frontier-orbital measurements that test it on the oligothiophenes $T_1$–$T_8$ (9–58 atoms), the longitudinal polarizability, a real-space charge-transfer decomposition, and the basis-set ceiling of the optical-gap reference. All model quantities are from the released checkpoint.

## S4.1 Size-scaling mechanism

**The pooling bound.** A read-out that sums local atomic contributions,

$$P_{\text{pool}} = \sum_{i=1}^{N} g(\mathcal{N}_i), \tag{1}$$

depends only on environments $\mathcal{N}_i$ within the network's receptive field $R_c$ (for message passing, the number of interaction layers times the radial cutoff: $4 \times 6 = 24$ Å here). Once the chain exceeds $R_c$, every interior unit shares the same bulk environment, so $P_{\text{pool}} = N\, g_{\text{bulk}} + g_{\text{edge}}$: a sum read-out is strictly extensive ($P \propto N$, exponent $\tau = 1$) and a mean read-out strictly intensive ($\tau = 0$). For *any* learned $g$ the per-unit response is frozen beyond $R_c$, so a pooled read-out cannot represent a super-linear ($\tau > 1$) collective response. The charge-weighted dipole read-outs used in practice, $\boldsymbol{\mu} = \sum_i (q_i \mathbf{r}_i + \boldsymbol{\mu}_i)$ with locally predicted charges $q_i$ [2, 3], do not evade it: the polarizability needs the field response $\partial q_i/\partial F$, which for a finite-cutoff model is local, and charge conservation ($\sum_i \partial q_i/\partial F = 0$) forces the bulk response to cancel and leaves a net response only within $R_c$ of the chain ends, so the charge-transfer polarizability is at most linear. Only genuinely non-local read-outs (global attention [4], charge equilibration [5]) escape the bound, but must then *learn* the size dependence from data and do not reliably extrapolate it.

**The Hamiltonian read-out.** The Hamiltonian read-out instead *computes* the size dependence. The polarizability is the linear response of the eigenpairs $\{\varepsilon_n, \psi_n\}$ of $\mathbf{H} = \mathbf{H}_{\text{DFT}} + \Delta\mathbf{H}_\theta$; retaining the dominant frontier (HOMO→LUMO) term of the sum-over-states response,

$$\alpha_{xx} \;\approx\; \frac{4\,|\langle \psi_H|\hat{x}|\psi_L\rangle|^2}{\varepsilon_L - \varepsilon_H}. \tag{2}$$

The correction $\Delta\mathbf{H}_\theta$ is local, but diagonalizing $\mathbf{H}$ is global: a local Hamiltonian has delocalized eigenstates (as a nearest-neighbour tight-binding chain has Bloch states), so the frontier orbitals spread over the conjugation length $\xi$, which is not bounded by $R_c$. Evaluated on these delocalized states with the extensive operator $\hat{x}$, both factors of Eq. (2) keep evolving with chain length well beyond $R_c$. Neither freezes at $R_c$, and the controlling length is $\xi$, set by the global eigenproblem, not $R_c$.

**Measured on the predicted Hamiltonian.** Figure S2 tests each step on the oligothiophenes. The longitudinal polarizability (panel a), the component of main-text Fig. 5a in which the super-linear response is largest, rises ∼19-fold from $T_1$ to $T_8$ (versus ∼12-fold for $\alpha_{\text{iso}}$); HARP tracks finite-field CCSD/cc-pVDZ across the whole reference range (signed error within 2.3% out to the $T_6$ limit), whereas the direct read-out ablation over-polarizes by a margin that grows to 47% at $T_6$, drifting toward the BP86 baseline. The ablation shares the backbone architecture and dataset of HARP (its training set omits the 2001–3000 index block, ∼2% of the data) but corrects the BP86 rather than the B3LYP baseline, at $l_{\max} = 3$. The baseline is not what makes it drift: the BP86-baseline Hamiltonian-read-out model of the reference ablation (Section S3) stays within +5.8% of CCSD in $\alpha_{xx}$ (+4.3% in $\alpha_{\text{iso}}$) through $T_5$, where the direct read-out is off by +35% (+21%), so the divergence follows the read-out. The frontier orbitals of the predicted Hamiltonian (panel b) are delocalized over the whole chain and keep evolving past the 24 Å receptive field (the chain is longer than 24 Å from $T_6$ on): the HOMO spread $\sigma_{\text{HOMO}} = \sqrt{\langle \hat{x}^2\rangle - \langle \hat{x}\rangle^2}$ grows nearly linearly with length ($\sim N^{0.92}$), and the transition dipole $|\langle \psi_H|\hat{x}|\psi_L\rangle|$ grows sub-linearly ($\sim N^{0.68}$, bending toward saturation as $L \to \xi$), while the frontier gap closes as $\sim N^{-0.38}$ (panel c). By Eq. (2) these

give $\tau \approx 2(0.68) + 0.38 = 1.75$. Panel d evaluates the two-level estimate directly: it scales as $N^{1.75}$, close to the $N^{1.58}$ of the full predicted $\alpha_{xx}$, and accounts for a nearly size-independent 67–75% of it beyond $T_1$, so the frontier term carries the size dependence. The exponent $\tau \approx 1.6$ is super-linear and lies between the pooled-read-out bound ($\tau = 1$) and the fully delocalized limit (a transition dipole $\propto L$ with a saturated gap would give $\tau = 2$): these chains are in the crossover $L \lesssim \xi$. The conclusion does not depend on the exact exponent: $\tau > 1$ is structurally impossible for the pooled read-out of Eq. (1).

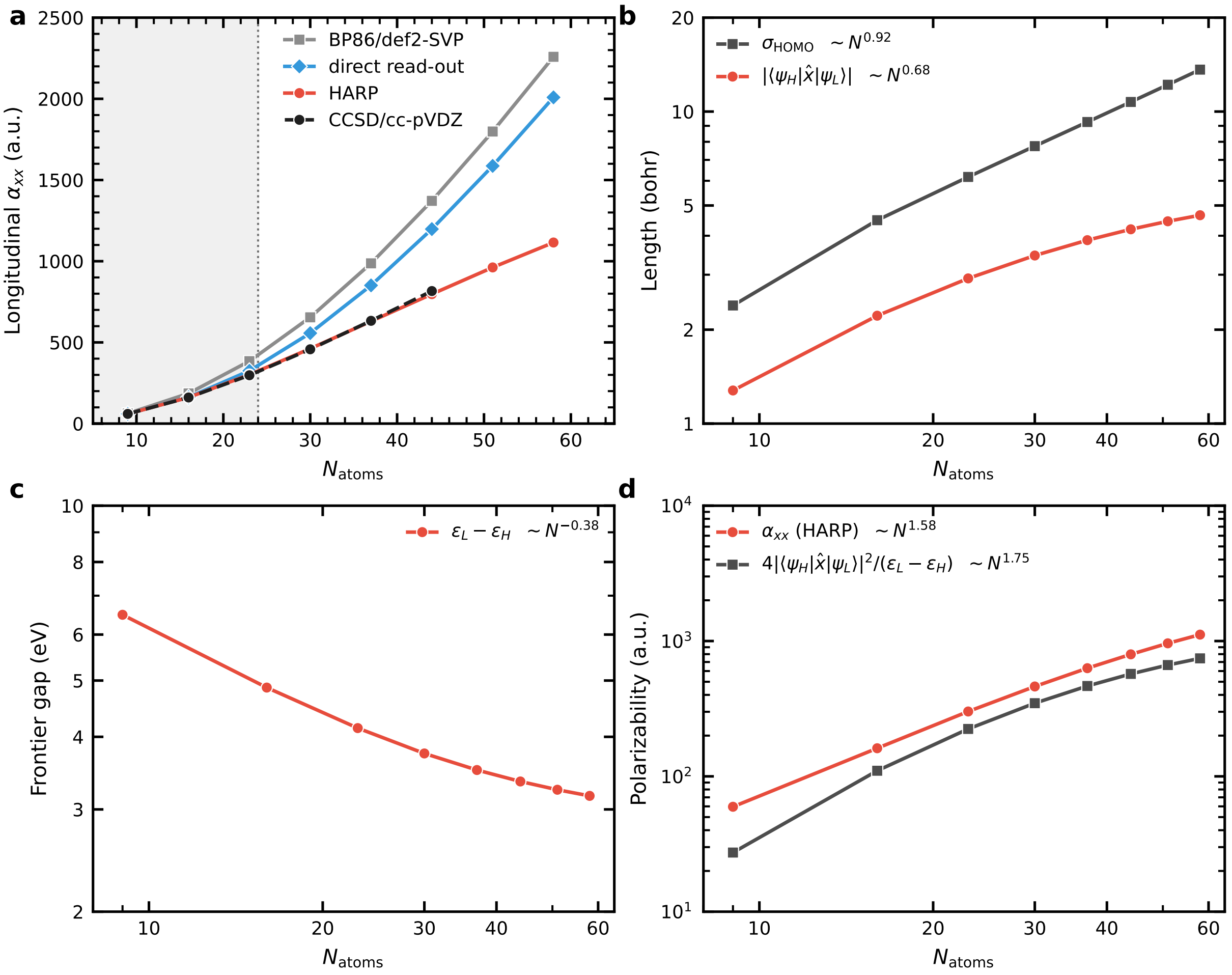


Figure S2: **The size dependence is computed by the predicted Hamiltonian's global eigenproblem.** Oligothiophenes $T_1$–$T_8$; model quantities from the released HARP checkpoint. **a** Longitudinal polarizability $\alpha_{xx}$, the companion to the isotropic $\alpha_{\text{iso}}$ of main-text Fig. 5a: HARP (red) tracks finite-field CCSD/cc-pVDZ (black, dashed; affordable to $T_6$), while the direct read-out ablation (blue) over-polarizes toward the BP86/def2-SVP baseline (grey). Shading and the dotted line mark the training size range ($\leq$ 24 atoms). **b** Frontier-orbital lengths of the predicted Hamiltonian, log–log: HOMO spread $\sigma_{\text{HOMO}}$ and HOMO→LUMO transition dipole $|\langle\psi_H|\hat{x}|\psi_L\rangle|$. **c** Frontier eigenvalue gap $\varepsilon_L - \varepsilon_H$. **d** Two-level estimate $4|\langle\psi_H|\hat{x}|\psi_L\rangle|^2/(\varepsilon_L - \varepsilon_H)$ (Eq. 2) against the full predicted $\alpha_{xx}$. Legend exponents are least-squares log–log slopes over $T_1$–$T_8$ (effective exponents over this size range).

### S4.2 Real-space picture: charge-transfer decomposition

The real-space mechanism behind the super-linear polarizability is visible in a distributed-polarizability analysis (Fig. S3): a Hirshfeld partition of the finite-field response density into a per-atom local dipole term and an inter-atomic charge-transfer term ($\Delta q \cdot R$) [6, 7] shows that the longitudinal polarizability of these chains is dominated by charge transfer. The decomposition requires the field-perturbed density, which we take from BP86/def2-SVP (finite field via the ORCA ELPROP module); BP86 over-polarizes these chains (Fig. S2a), so the decomposition is a qualitative, real-space picture rather than a quantitative reference. The local atomic-dipole contribution is small and grows *linearly* in ring count (a near-constant 8.3–9.2 a.u. per added ring; the effective log–log slope is below one only because of the end-group offset), whereas the charge-transfer contribution is super-linear ($\sim N^{2.1}$) and accounts for 68% of $\alpha_{xx}$ at $T_1$, rising to 93% at $T_4$ and 97% at $T_8$. The super-linear growth is thus almost entirely a collective, delocalized charge-flow effect—precisely the contribution that a pooling or induced-dipole read-out, restricted to per-atom terms summed or averaged over the graph, structurally cannot represent, and that a small-molecule training set cannot teach. The Hamiltonian read-out captures it because the polarizability is computed as the linear response of the predicted Hamiltonian (a global sum over states through the delocalized molecular orbitals of $\mathbf{H}$), so the collective charge flow emerges from the same quantum mechanics that produces it in the reference, even though the learned correction $\Delta\mathbf{H}_\theta$ is strictly local.

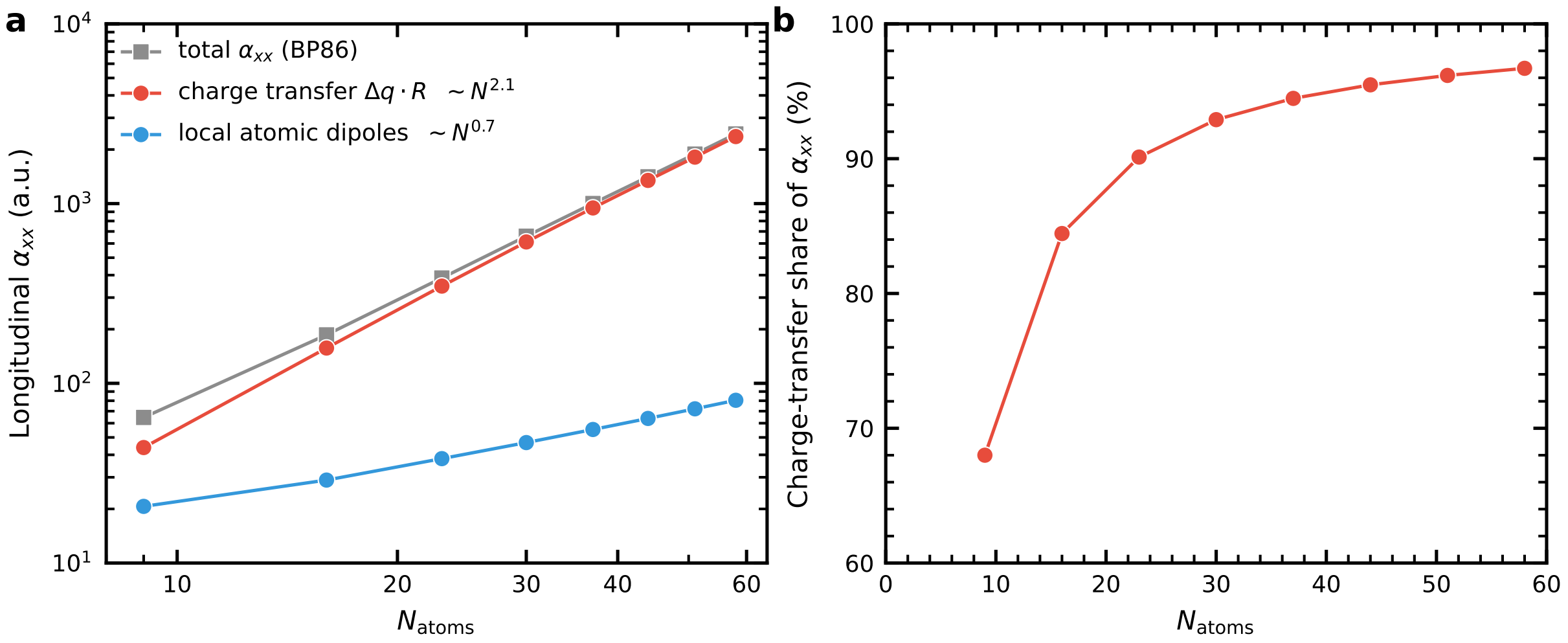


Figure S3: **Conjugated-chain polarizability is dominated by inter-atomic charge transfer.** Distributed-polarizability (Hirshfeld) decomposition of the BP86/def2-SVP finite-field longitudinal polarizability of the oligothiophenes $T_1$–$T_8$. **a** Total $\alpha_{xx}$ (grey) and its two parts, log–log: the per-atom local dipole response (blue) and the inter-atomic charge-transfer term $\Delta q \cdot R$ (red), with least-squares slopes. **b** Charge-transfer share of $\alpha_{xx}$, rising from 68% ($T_1$) to 97% ($T_8$). Pooling and induced-dipole read-outs can represent the local term but not the collective charge flow.

### S4.3 Basis-set ceiling of the optical-gap reference

The optical-gap label is the lowest EOM-CCSD/cc-pVDZ vertical excitation (main text Methods). Both the model and this reference lie ~1 eV above experimental UV–Vis maxima for the oligothiophenes; we attribute the offset to the reference level, not the surrogate. Table S10 isolates the basis-set contribution by recomputing the EOM-CCSD excitation with diffuse functions (aug-cc-

pVDZ) on the same geometries. Adding diffuse functions lowers the excitation by 0.41 eV at $T_1$ and 0.31 eV at $T_2$, recovering a large fraction of the offset toward experiment. The remainder is ascribable to further basis incompleteness (toward the complete-basis limit), the absence of triples in EOM-CCSD, and the vertical-versus-onset and gas-versus-solution differences between a computed vertical excitation and a measured solution absorption maximum.

Full coupled-cluster triples for excitations (CC3 or EOM-CCSDT) would quantify the method (triples) contribution directly, but scale as $\mathcal{O}(N^{7-8})$ and are intractable for systems of this size; we therefore bound the triples contribution with literature benchmarks for thiophene and short oligothiophenes [8, 9] rather than computing it explicitly. This is the binding accuracy limitation discussed in the main text, and the motivation for the fundamental-gap (IP−EA) and diffuse-basis dataset roadmap.

Table S10: Lowest singlet vertical excitation (eV) of oligothiophenes: the model, its EOM-CCSD/cc-pVDZ label, the same reference with diffuse functions (aug-cc-pVDZ), and the experimental UV–Vis absorption maximum.

| | this work | EOM-CCSD/cc-pVDZ | EOM-CCSD/aug-cc-pVDZ | exp. ($\lambda_{\max}$) |
|---|---|---|---|---|
| $T_1$ | 6.49 | 6.52 | 6.11 | 5.37 |
| $T_2$ | 4.87 | 4.88 | 4.57 | 4.10 |

# S5 Computational efficiency

## S5.1 Inference cost: full timing analysis

The neural inference of the released model—network forward pass, assembly and diagonalization of **H**, the screened response, and evaluation of all properties—was timed per molecule on a single NVIDIA A100 (40 GB) (protocol below). Figure S4 places it on the same axes as the four reference electronic-structure methods. It takes 0.10–0.14 s (median 0.126 s, of which ∼0.10 s is the network forward pass) and is nearly flat over the 5–24-atom test range ($\sim N^{0.24}$): at batch size 1 it is dominated by GPU kernel-launch latency rather than floating-point work. The B3LYP/def2-SVP baseline takes a median 21 s on the same molecules, so the neural part is a ∼165-fold smaller contribution ($< 1\%$ of the total) and the inference cost is that of the DFT single point. Because the neural cost grows far more slowly with $N$ than any electronic-structure method, the speedup is a lower bound that widens with system size.

The extrapolation of Fig. S4 is borne out by direct measurement on the oligothiophene series itself (the systems of main-text Fig. 5). Figure S5 times a single point of each reference method on $T_1$–$T_6$ (9–44 atoms) at a fixed 16 MPI ranks on one node. The CCSD finite-field polarizability grows from 17 s at $T_1$ to 5.7 h per field point at $T_6$ (seven points for the full tensor), and the EOM-CCSD gap reaches 3.3 h at $T_4$, whereas the B3LYP/def2-SVP baseline that sets the MEHnet cost stays between 10 and 41 s across the same range. The measured CCSD-to-baseline ratio therefore widens from ∼90× at 30 atoms to ∼500× at 44 atoms, confirming on the actual large systems the size-growing speedup anticipated from the smaller test set.

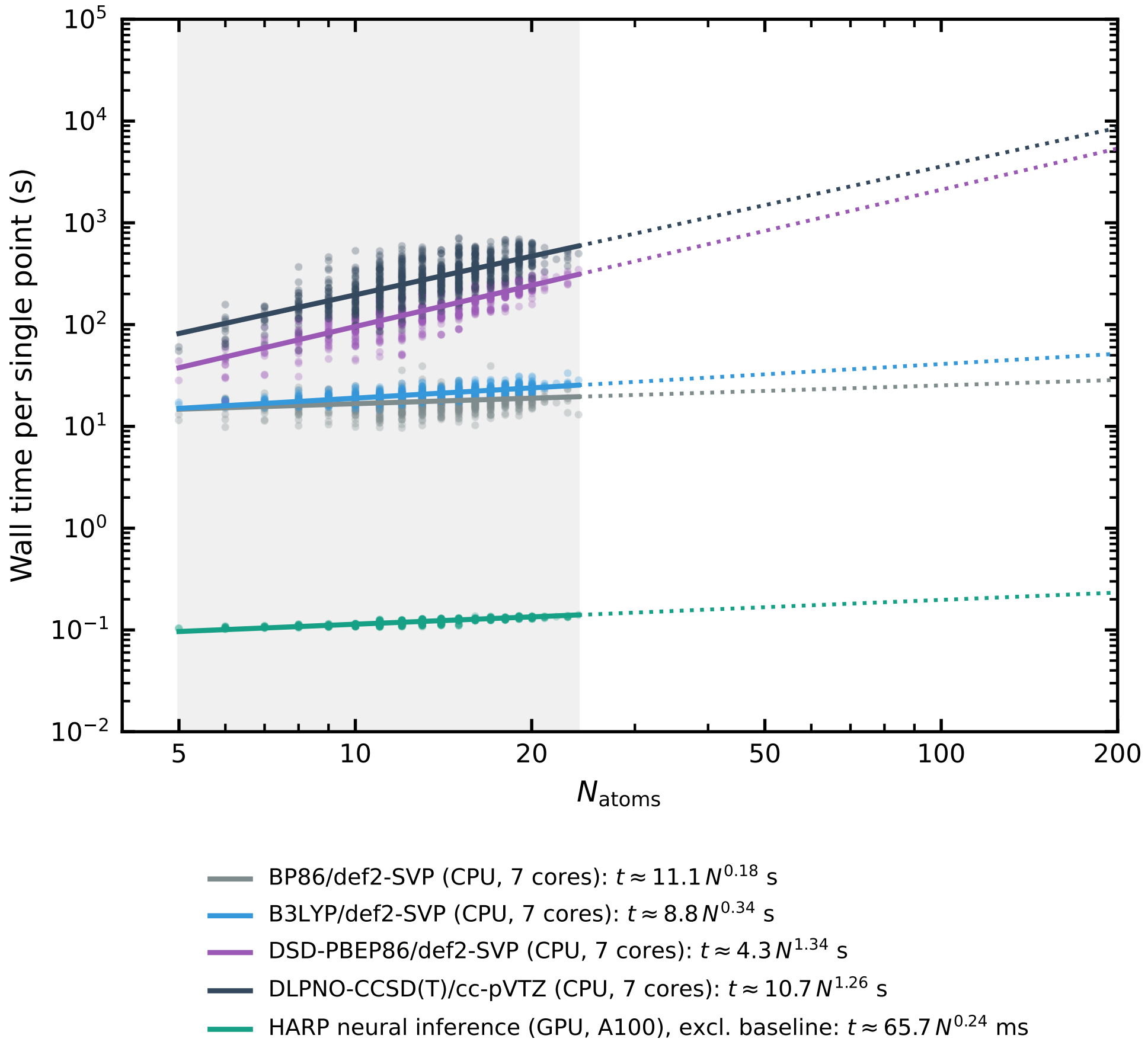


Figure S4: **The neural inference is a negligible addition to the DFT baseline.** Per-molecule wall time on the 959 test molecules: HARP neural inference (GPU, green; excludes the baseline) and the electronic-structure methods (CPU, 7 cores; ORCA TOTAL RUN TIME at dataset generation). Lines are power-law fits, solid over the measured range (shaded, 5–24 atoms) and dotted where extrapolated to $N = 200$ atoms. The full HARP cost is the green curve plus the B3LYP/def2-SVP single point (blue).

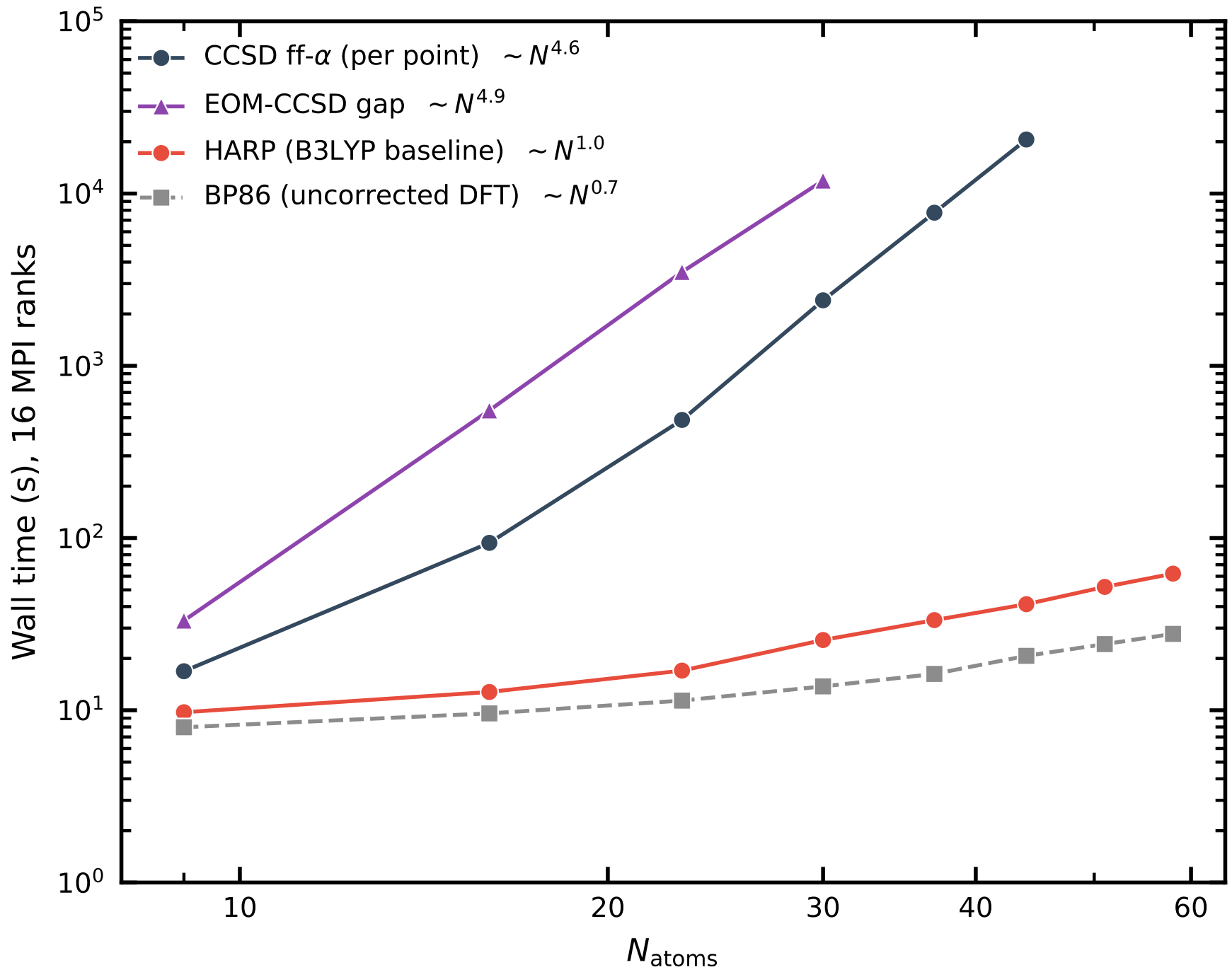


Figure S5: **Coupled-cluster cost on the oligothiophenes.** Measured single-point wall time (ORCA TOTAL RUN TIME, 16 MPI ranks, one node) of the reference electronic-structure methods on the oligothiophenes $T_1$–$T_6$: CCSD/cc-pVDZ finite-field $\alpha$ (per field point), EOM-CCSD/cc-pVDZ optical gap, and the B3LYP/def2-SVP baseline that dominates the HARP inference cost (the neural inference adds ∼0.1 s, Fig. S4); BP86/def2-SVP shown for reference. Fitted power laws (labels) are effective slopes over this size range; the coupled-cluster methods run 2–3 orders of magnitude slower than the DFT baseline, with the gap widening as $N$ grows (CCSD/B3LYP $\approx 94\times$ at 30 atoms and $\approx 499\times$ at 44 atoms).

### S5.2 Neural-network timing protocol

Per-molecule inference on a single NVIDIA A100 (40 GB), batch size 1, `eval` mode under `no_grad`, 5 warmup passes, minimum of 3 timed passes bracketed by CUDA synchronization. The timed interval covers the network forward pass, the Hamiltonian eigensolver and screened response, and the evaluation of every property; graph construction and data loading are excluded.

## S6 Architecture comparison: MEHnet vs. HARP

The present model (HARP) shares its central design principle with the original multi-task Hamiltonian-learning model of Tang *et al.* [10] (MEHnet): rather than reading each property out of pooled atomic features, both predict a local correction to a baseline one-electron (Kohn–Sham) Hamiltonian and then evaluate every observable from that Hamiltonian by its defining quantum-mechanical operator. This shared inductive bias is what gives both models their size transferability. What changed in HARP is the machinery around that read-out: the equivariant backbone, the chemical scope, the choice of baseline (and the gap treatment it makes possible), the training-stability scheme, and the dataset. Table S11 summarizes the changes; we discuss the consequential ones below.

**Backbone.** MEHnet used a two-layer EGNN in which equivariant node and edge features are built from scalar/vector messages, with angular momenta up to $l_{\max} = 2$. HARP replaces it with EquiformerV2, a deeper (4-interaction-layer) spherical-harmonic graph-attention network in the eSCN lineage that carries features up to $l_{\max} = 4$—the angular resolution required to represent the $d \times d$ on-site Fock blocks that the third-row elements introduce—and uses attention rather than fixed convolution. The richer angular resolution and greater depth are what make the single-Hamiltonian read-out accurate across nine elements with their more varied bonding (lone pairs, hypervalency, heteroatom polarity) rather than the C/H bonding of MEHnet.

**Baseline and gap.** MEHnet built its correction on top of a BP86 (local GGA) baseline. A GGA Kohn–Sham gap underestimates the optical gap with a slope well below unity, so MEHnet had to learn a two-parameter linear rescaling of the HOMO–LUMO gap, $E_g = (1 + G_1)(\varepsilon_{\mathrm{L}} - \varepsilon_{\mathrm{H}}) + G_2$, with $G_1, G_2$ produced by an attention-pooled MLP head. HARP instead builds on a B3LYP hybrid baseline, whose Kohn–Sham gap already tracks the correlated optical gap with near-unit slope (Section S3); the gap is then read directly as the frontier eigenvalue difference of $\mathbf{H}$, and the learned rescaling is dropped entirely. This removes a property-specific head that sat outside the Hamiltonian read-out—and not an innocuous one: the $G_1, G_2$ correctors were attention-pooled learned scalars, i.e. exactly the pooling-type quantity whose size behaviour the read-out argument says is not guaranteed. With them gone, the gap is obtained by the same operator-on-$\mathbf{H}$ route as every other property, which is both cleaner and extends the size-transferability argument to the full property suite; the out-of-distribution gap of main-text Fig. 5b is the direct test. The reference ablation of Section S3 confirms this is the only property for which the baseline choice materially matters.

**Energy head and response anchoring.** MEHnet read the total energy as the occupied-eigenvalue sum and learned its screening matrix from scratch. HARP makes both channels Δ-learning quantities: an explicit zero-initialized per-atom energy head $\Delta E_\theta$, and a screening matrix $\mathbf{T} = \mathbf{T}_{\mathrm{DFT}} + \Delta\mathbf{T}_\theta$ whose baseline part is constructed from the baseline response, $\mathbf{T}_{\mathrm{DFT}} = \boldsymbol{\alpha}_{\mathrm{DFT}}^{-1} - \boldsymbol{\alpha}^0(\mathbf{H}_{\mathrm{DFT}})^{-1}$, where $\boldsymbol{\alpha}_{\mathrm{DFT}}$ is the baseline linear-response polarizability and $\boldsymbol{\alpha}^0$ the bare

Table S11: Architecture changes from MEHnet [10] to the present work (HARP). Both models share the effective-Hamiltonian read-out $\mathbf{H} = \mathbf{H}_{\text{baseline}} + \Delta\mathbf{H}_\theta$ and the linear-response polarizability; the rows below are where HARP differs.

| | **MEHnet** (Tang *et al.* 2024) | **HARP** (this work) |
|---|---|---|
| Backbone | E(3)-equivariant GNN (EGNN), 2 graph-convolution layers, $l_{\max} = 2$ | SO(3)-equivariant spherical-harmonic graph-attention network (EquiformerV2 [11], eSCN lineage [12]); 4 interaction layers, $l_{\max} = m_{\max} = 4$ (required by the $d \times d$ Fock blocks of Si–Cl), 64 channels, 8 heads ($\sim$5.6 M params) |
| Element scope | Hydrocarbons (C, H) | Nine main-group elements (H, C, N, O, F, Si, P, S, Cl) |
| Baseline | BP86/cc-pVDZ (local GGA) with a many-body energy shift | B3LYP/def2-SVP (hybrid) |
| Correction heads | $\Delta\mathbf{H}_\theta$ plus a screening matrix learned from scratch; energy read as the occupied-eigenvalue sum $2\sum_i \varepsilon_i$ | Zero-initialized $\Delta E_\theta$, $\Delta\mathbf{H}_\theta$, $\Delta\mathbf{T}_\theta$; explicit energy head, and the screening matrix anchored to the baseline response, $\mathbf{T} = \mathbf{T}_{\text{DFT}} + \Delta\mathbf{T}_\theta$ ($\Delta = 0$ reproduces the baseline $\alpha$ exactly) |
| Gap read-out | Learned linear rescaling of the HOMO–LUMO gap, $E_g = (1 + G_1)(\varepsilon_{\text{L}} - \varepsilon_{\text{H}}) + G_2$, with $G_1, G_2$ from attention-pooled scalars | Frontier eigenvalue difference directly; *no* learned rescaling ($G_1, G_2$ dropped) |
| Training stability | Perturbation-theory back-propagation through the eigensolver for near-degenerate levels | Diabatic level-tracking (occupied subspace by maximal overlap with the baseline density), complementary to the degeneracy treatment; needed only on unfiltered data—the released model trains without it (Methods) |
| Training data | $\sim$10,500 MD configurations of 85 hydrocarbons (MW $<$ 100) | $\sim$42k quality-filtered PubChem molecules spanning the nine elements |

sum-over-states response of the baseline Fock matrix. At $\Delta\mathbf{T}_\theta = 0$ the screened read-out therefore reproduces the baseline DFT polarizability exactly, extending the correct-by-construction $\Delta$-learning limit from the energy and Hamiltonian channels to the response channel.

**Training stability.** Both models must back-propagate through an eigensolver whose gradients are ill-conditioned near degeneracies. MEHnet stabilized this with a perturbation-theory expansion of the eigenvalue/eigenvector response. HARP adds a diabatic *level-tracking* scheme: the occupied subspace at each step is identified by maximal overlap with the baseline occupied density rather than by strict energy ordering, so the occupied/virtual assignment varies continuously through a level crossing. This addresses a failure mode that becomes more prevalent across the chemically diverse, heteroatom-containing molecules of the present dataset, where the learned correction can reorder frontier orbitals during training.

Table S12: Per-property accuracy of MEHnet [10] and HARP. *The two columns are not directly comparable* — MEHnet is in-domain RMSE against CCSD(T)/cc-pVDZ on 85 hydrocarbons, HARP is MAE against composite CCSD(T)/cc-pVTZ on 959 nine-element molecules (see text). They are shown together only to indicate that HARP retains or improves per-property accuracy while covering a far broader chemical space. The dipole and quadrupole rows in particular are dominated by the difference in test distribution, not by model quality (see text). For comparability with MEHnet, the dipole and quadrupole rows here are full-vector magnitudes (dipole in Debye, quadrupole Frobenius norm in a.u.), distinct from the per-component a.u. errors of Table S5.

| Property | Unit | **MEHnet** (ID RMSE) | **HARP** (MAE) |
|---|---|---|---|
| Energy | $\mathrm{kcal\,mol^{-1}}$ | $1.38^{\dagger}$ | 0.23 |
| Gap (1st exc.) | eV | 0.26 | 0.067 |
| Dipole | D | 0.006 | 0.022 |
| Quadrupole | a.u. | 0.03 | 0.044 |
| Mulliken charge | $e$ | 0.04 | 0.0053 |
| Mayer bond order | — | 0.02 | 0.0066 |
| Polarizability | a.u. | 1.85 | 0.121 |

$^{\dagger}$Rescaled to per-molecule from MEHnet's reported 0.11 $\mathrm{kcal\,mol^{-1}}$ per atom using the config-weighted mean of 12.6 atoms per molecule.

**Scope and data.** MEHnet was trained and validated on ∼10,500 MD configurations of 85 hydrocarbons (molecular weight < 100) and demonstrated on aromatic and polymeric hydrocarbons. HARP is trained on ∼42k quality-filtered PubChem molecules spanning all nine elements. The combination of the more expressive backbone, the hybrid baseline, and the larger and broader dataset is what extends the effective-Hamiltonian approach from a hydrocarbon proof of concept to a model that covers the main group while retaining the out-of-distribution size scaling that motivates the architecture.

**Accuracy.** The two models cannot be compared by their headline error numbers directly, for three reasons. (i) *Metric*: MEHnet reported root-mean-square error (RMSE), whereas the HARP numbers (main-text Table 1; Table S5) are mean absolute errors (MAE), which are systematically smaller. (ii) *Reference level*: MEHnet was benchmarked against CCSD(T)/cc-pVDZ (EOM-CCSD and CCSD/cc-pVDZ for the gap and polarizability), whereas HARP uses the larger, closer-to-complete composite CCSD(T)/cc-pVTZ reference (main-text Methods), a harder target. (iii) *Test distribution*: MEHnet's test set is molecular-dynamics configurations of 85 hydrocarbons, while HARP's is 959 molecules spanning all nine elements. To make the energy row commensurable we have rescaled MEHnet's reported per-atom energy error to a per-molecule value using the config-weighted mean of 12.6 atoms per molecule over its 85-molecule test set. With those caveats stated, Table S12 places the two side by side; HARP holds or improves on MEHnet across most properties while being evaluated on a substantially harder and more chemically diverse benchmark.

The dipole and quadrupole rows deserve a caveat beyond those above: their absolute errors are set largely by the test distribution rather than by model quality. MEHnet's hydrocarbon test molecules are nearly nonpolar, so their reference dipoles and quadrupoles — and hence the achievable absolute errors — are intrinsically small. HARP's test set contains strongly polar species

(N, O, F, Cl bonds), whose multipoles are an order of magnitude larger, so the larger absolute multipole error does *not* indicate a less accurate model — it reflects a harder, more polar target. A fair multipole comparison would require evaluating both models on a common, chemically matched test set, which is outside the scope of this work.

# S7 Scope map of released ML models: charge, spin, and geometry regime

Main-text Table 2 asks *which properties* each released model can deliver on this benchmark's chemistry. Table S13 asks the complementary question: over *which chemical states and geometries* is each model trained or validated—charged species, open-shell/spin states, off-equilibrium geometries, and transition states or otherwise reactive structures. Every cell was checked against the cited paper or repository (July 2026). Two systematic distinctions run through the table. First, the machine-learned functionals (DM21, Skala) inherit arbitrary charge, spin, and geometry handling from the Kohn–Sham SCF machinery they plug into, so their marks reflect what their *training and validation* actually covered. Second, among the neural models only SpookyNet and UMA condition on total charge and spin as explicit inputs; QHNet, HELM, MACE-OFF23, and HARP read only element types and positions, and therefore cannot distinguish charge states or multiplicities of the same geometry.

HARP's row is deliberately narrow and matches the dataset scope stated in the main text: neutral, closed-shell singlets at near-equilibrium PubChem conformers. Extending the same effective-Hamiltonian machinery to charged and open-shell species is the natural next step—the dataset already stores bt-PNO-IP/EA-EOM-CCSD ionization potentials and electron affinities for every molecule (Section S1.2)—and is part of the roadmap discussed in the main text.

Table S13: Chemical-state and geometry scope of the models of main-text Table 2, per their training/validation data. ✓ covered in training or validated in the cited work; ✓ partial (see notes); – not covered. Notes: DM21 trains on ionization/electron-affinity reactions and fractional charge *and* spin constraints, and validates bond breaking and (diradical) barrier heights. Skala trains on IP/EA data, distorted structures, and BH9 reaction barriers; its open-shell coverage is atomic (partial). SpookyNet and UMA take total charge and spin as inputs (SpookyNet: singlet/triplet $CH_2$, $Ag_3^{\pm}$; UMA: OMol25 variable charge/spin and reactive structures). QHNet's QH9 adds 300 K MD snapshots (thermal, not reactive). HELM's OMol_CSH training slice is closed-shell by construction (restricted Kohn–Sham) with unreported charge composition; its off-equilibrium mark refers to MD-snapshot and conformer benchmarks (MD17/QM7, $\nabla^2$DFT). MACE-OFF23's SPICE training set is neutral (ion pairs removed) with MD-sampled conformations.

| Method | Charged | Open-shell/spin | Off-equilibrium | TS/reactive |
|---|---|---|---|---|
| **HARP (this work)** | – | – | – | – |
| QHNet [13] | – | – | ✓ | – |
| HELM [14] | – | – | ✓ | – |
| DM21 [15] | ✓ | ✓ | ✓ | ✓ |
| Skala [16] | ✓ | ✓ | ✓ | ✓ |
| SpookyNet [4] | ✓ | ✓ | ✓ | – |
| MACE-OFF23 [3] | – | – | ✓ | – |
| UMA / OMol25 [17] | ✓ | ✓ | ✓ | ✓ |

# S8 Computational details and reference protocols

## S8.1 Geometry construction

Oligothiophenes were built as planar, $s$-trans, $\alpha, \alpha'$-linked chains with alternating sulfur orientation, using gas-phase microwave thiophene bond parameters (S–$C_\alpha$ 1.714 Å, $C_\alpha$–$C_\beta$ 1.370 Å, $C_\beta$–$C'_\beta$ 1.423 Å, inter-ring 1.448 Å, C–H 1.080 Å, ∠C–S–C 92.2°); the geometry generator is in the released code.

## S8.2 ORCA reference protocols for the oligothiophenes

The reference calculations on the oligothiophenes (main-text Fig. 5) used ORCA 6.0 [1]; the dataset protocol is given in Section S1.2. BP86 polarizabilities: analytic via the `%ELPROP Polar` module (def2-SVP). CCSD polarizabilities: finite field, $\alpha_{ii} = -[E(+F_i) + E(-F_i) - 2E_0]/F^2$ with $F = 5 \times 10^{-4}$ a.u. (cc-pVDZ). EOM-CCSD excitations: `%MDCI nroots 3` on an RHF reference (cc-pVDZ; aug-cc-pVDZ for the basis-ceiling study). Single-point HOMO–LUMO gaps: BP86 and B3LYP (def2-SVP). Parallel ORCA runs used OpenMPI 4.1.


\end{document}